 \documentclass[final,5p,times,twocolumn,authoryear]{elsarticle}

\usepackage{amssymb}
\usepackage{amsmath}
\usepackage{xcolor}
\usepackage[normalem]{ulem}
\usepackage[
  colorlinks=true,   
  linkcolor=blue,    
  citecolor=blue,    
  urlcolor=blue,     
  breaklinks=true    
]{hyperref}
\usepackage{xurl}

\journal{Earth Energy Science}

\begin{document}


\begin{frontmatter}



\title{Machine learning-based upscaling of rock
permeability from pore scale to core scale: effect of
training dataset size and sub-core volumes} 


\author[label1]{Yaotian Guo} 
\author[label1]{Fei Jiang}
\author[label2]{Takeshi Tsuji}
\author[label3]{Yoshitake Kato}
\author[label3]{Mai Shimokawara}
\author[label4]{Lionel Esteban}
\author[label4]{Mojtaba Seyyedi}
\author[label4]{Marina Pervukhina}
\author[label5]{Maxim Lebedev}
\author[label3]{Ryuta Kitamura}

\affiliation[label1]{%
  organization={Department of Mechanical Engineering, Yamaguchi University},
  addressline={2-16-1 Tokiwadai},
  city={Ube},
  postcode={755-8611},
  state={Yamaguchi},
  country={Japan}}

\affiliation[label2]{%
  organization={School of Engineering, The University of Tokyo},
  addressline={7-3-1 Hongo, Bunkyo-ku},
  city={Tokyo},
  postcode={113-8656},
  state={Tokyo},
  country={Japan}}

\affiliation[label3]{%
  organization={Japan Organization for Metals and Energy Security (JOGMEC)},
  addressline={1-2-2 Hamada, Mihama-ku},
  city={Chiba},
  postcode={261-0025},
  state={Chiba},
  country={Japan}}

\affiliation[label4]{%
  organization={CSIRO Energy Business Unit},
  addressline={26 Dick Perry Avenue, Kensington},
  city={Perth},
  postcode={6151},
  state={WA},
  country={Australia}}

\affiliation[label5]{%
  organization={Edith Cowan University},
  addressline={270 Joondalup Drive},
  city={Joondalup},
  postcode={6027},
  state={WA},
  country={Australia}}

\begin{abstract}
Permeability characterizes the capacity of porous formations to conduct fluids, thereby governing the performance of carbon capture, utilization, and storage (CCUS), hydrocarbon extraction, and subsurface energy storage. A reliable assessment of rock permeability is therefore essential for these applications.
Direct estimation of permeability from low-resolution CT images of large rock samples offers a rapid approach to obtain permeability data. However, the limited resolution fails to capture detailed pore-scale structural features, resulting in low prediction accuracy. To address this limitation, we propose a convolutional neural network (CNN)-based upscaling method that integrates high-precision pore-scale permeability information into core-scale, low-resolution CT images. In our workflow, the large core sample is partitioned into sub-core volumes, whose permeabilities are predicted using CNNs. The upscaled permeability at the core scale is then determined through a Darcy flow solver based on the predicted sub-core permeability map. Additionally, we examine the optimal sub-core volume size that balances computational efficiency and prediction accuracy. This framework effectively incorporates small-scale heterogeneity, enabling accurate permeability upscaling from micrometer-scale pores to centimeter-scale cores. 
\end{abstract}



\end{frontmatter}



\section{Introduction}
\label{sec:intro}
Reliable understanding of the physical properties of porous rocks is fundamental for advancing subsurface energy utilization, such as oil and gas recovery \citep{massarweh2022review,koroteev2021artificial}, geothermal energy exploitation \citep{rybach2010future,wu2024geothermal}, carbon capture and storage (CCS) \citep{bachu2008ccs,raza2019significant}, and underground hydrogen storage (UHS) \citep{bade2024hydrogen,taiwo2024storage}.
In this context, the key physical properties of porous rocks include porosity, permeability, formation factor, capillary pressure, pore size distribution, and pore connectivity. Among these, permeability is the most critical parameter for characterizing fluid transport, as it represents the material’s ability to allow fluid flow through its interconnected pore network. 
Permeability is highly dependent on the pore structure, and the heterogeneity of this structure often varies across a wide range of length scales, leading to the phenomenon of scale-dependent permeability \citep{bear2013dynamics,blunt2001flow}.
This scale dependency presents a significant challenge. Permeability estimated at the pore scale often differs substantially from measurements at the core or field scale due to the heterogeneity and anisotropy of the porous medium \citep{clauser1992permeability,esmaeilpour2021scale,esmaeilpour2023estimating}. 
To address this issue, upscaling methods are essential for improving large-scale permeability estimations based on small-scale data while incorporating details of heterogeneity. \citet{onimisi2023constrained} employed Proper Orthogonal Decomposition (POD) and Nonnegative Proper Orthogonal Decomposition (NPOD) models, coupled with interpolation algorithms, to upscale permeability from fine-scale reservoir models to coarse-scale representations while maintaining geological consistency. Similarly, \citet{elmorsy2023rapid} integrated a novel analytical approach with physics-informed neural networks (PINNs) to achieve rapid permeability upscaling from pore to core scale in digital porous media.
In a related vein, \citet{yousefzadeh2023fast} developed a hybrid deep learning framework that combined Convolutional Long-Short-Term Memory (ConvLSTM) and particle swarm optimization, leveraging the fast marching method to upscale permeability from pore-scale to core-scale systems. \citet{mishra2024pore} utilized graph theory-based techniques to upscale permeability from dynamic pore-scale models to Darcy-scale representations, emphasizing the preservation of flow properties. Moreover, \citet{jiang2024svm} incorporated Minkowski functionals for rock type classification, employing Support Vector Machines (SVM) and Gaussian Mixture Models (GMM) to upscale permeability from 3D pore-scale models to larger-scale rock representations.

Despite recent progress, accurately predicting permeability across a wide range of rock scales requires the integration of high-precision permeability data obtained at small scales, which effectively captures local heterogeneity, into larger-scale models. Especially, estimating large-scale permeability from low-resolution rock CT images poses a significant challenge, primarily due to the inability of low-resolution images to adequately capture the effects of small-scale heterogeneity. The blurring of narrow pore throats in low-resolution images hampers precise pore structure analysis, resulting in inaccuracies in permeability predictions at pore scale. Therefore, high-resolution CT images are indispensable to calculate reliable permeability data that includes detailed information on small-scale heterogeneity. High-resolution CT images enable the estimation of highly accurate permeability data through direct simulations using methods such as the Lattice Boltzmann Method (LBM). The LBM is particularly effective for modeling complex geometries and multiphase flows with high precision \citep{tsuji2016characterization, jiang2014changes, jiang2015impact, jiang2017estimation,wang2023recent, suss2023hybrid, li2016lattice}. However, although high-resolution CT images facilitate precise calculations of rock properties, their coverage is limited to small representative volumes, which is especially restrictive for highly homogeneous rocks. 
In contrast, low-resolution CT images provide coverage over larger sample volumes but lack the accuracy needed for calculating pore-scale physical properties. Therefore, accurate permeability upscaling requires integrating high-precision data from high-resolution CT images with broader-scale data from low-resolution CT images at the core scale. Therefore, it is essential to develop methods that utilize this integrated dataset to predict permeability across a wider range of scales.

Recently, deep learning techniques have been applied to the analysis of rock CT images. In particular, it has been demonstrated that this technology is effective in extracting the geometric properties of pores \citep{wang2021deep,roslin2022processing,hayatdavoudi2025comparative}.
Furthermore, deep learning techniques are increasingly used to predict characteristics such as 3D flow fields \citep{ko2023prediction}, capillary pressure \citep{telvari2023prediction}, permeability \citep{chen2023predicting,wang2023image,khan2024machine,kang2024hybrid}, and relative permeability \citep{telvari2023prediction,xie2023direct,kalule4781996relative} using 3D digital rock models.  \citet{telvari2023prediction} utilized a 3D CNN to predict relative permeability and capillary pressure from 3D images of digital sandstones. \citet{ko2023prediction} used a U-Net architecture CNN to predict the 3D velocity field and permeability from 3D binary images of reticulated foam. In \citet{chen2023predicting}, the authors utilized CNN to predict nine 3D physical properties, including permeability, from a single 2D slice of porous rocks. \citet{xie2023direct} used deep learning models (CNN, ConvLSTM-FC, ConvLSTM-CNN) to predict relative permeability curves by using 3D digital rock images and fluid/rock physical properties (such as wettability and interfacial tension) as inputs.
\citet{kalule4781996relative} combined physics-informed machine learning and reinforcement learning to predict relative permeability and capillary pressure.

Recognizing the capabilities of deep learning techniques, \citet{jiang2023upscaling} proposed a workflow that leverages deep learning in combination with digital rock physics to integrate multi-scale data, bridging low-resolution CT images with high-resolution physical property calculations. In this workflow, small subvolumes were extracted from high-resolution CT images, and their permeabilities were determined using the LBM. These permeability values, paired with the corresponding low-resolution CT images, were utilized to train a CNN. The trained CNN was then employed to predict the permeability map at the core scale, and the upscaled permeability of the entire core was subsequently calculated using a Darcy flow solver. However, the optimal subvolume size for this workflow remains uncertain. 
This study seeks to further validate the effectiveness of  \citet{jiang2023upscaling}’s workflow.
The study also evaluates the required number of subvolumes needed to train the CNN to achieve acceptable prediction accuracy. Ultimately, permeability upscaling from the millimeter sub-core scale to the centimeter large core sample was successfully achieved.

\section{Rock samples}
\begin{figure}[t]
\centering
\includegraphics[width=\columnwidth]{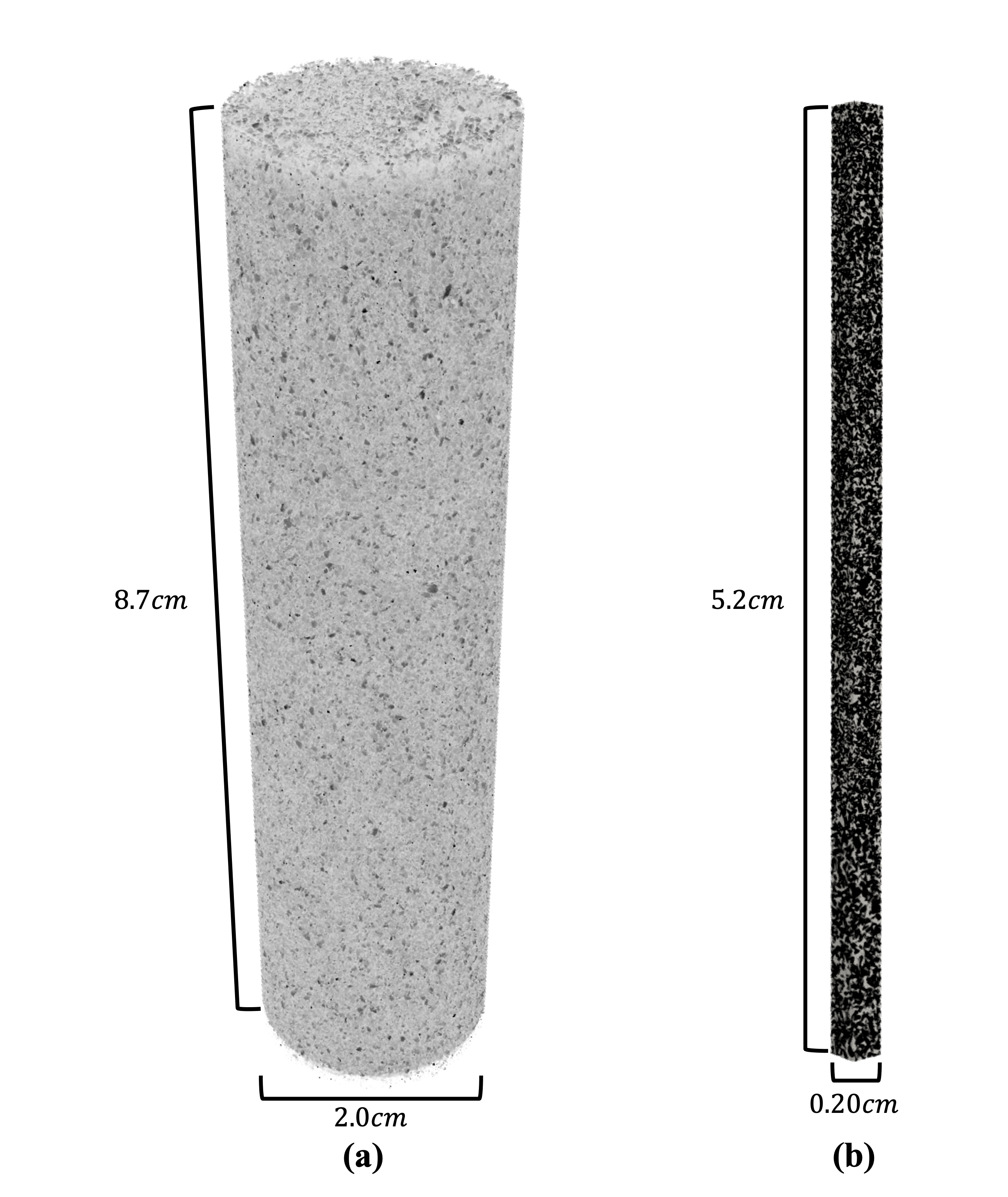}
\caption{Rock samples scanned with two different resolutions: (a) low-resolution for entire rock core, (b) high-resolution for center core region.}
\label{fig:rock}
\end{figure}

A Boise sandstone core was used in this study because its grain-size distribution, pore structure, and petrophysical characteristics have been extensively characterized, making it an ideal candidate for permeability upscaling modeling~\citep{louis2024role,cheung2012effect,li2021calculation}. 
It typically consists of approximately 40\% quartz and 50\% feldspars, and exhibits a porosity of about 24\%. The pore sizes are mostly in the range of about 50~$\mu$m, and the dominant grain size is around 100~$\mu$m, as measured from thin section optical images~\citep{jiang2023upscaling}.
Detailed information regarding the physical properties of the same type of Boise sandstone is available in our earlier publication~\citep{jiang2023upscaling}.
The micro X-ray CT imaging of the samples was conducted using a micro-CT scanner (Xradia Versa XRM-500) at two resolutions: 20$\mu m$/voxel (coarse) and 3$\mu m$/voxel (fine). For the coarse resolution, the complete rock sample at the centimeter scale was scanned, covering a diameter of 2.0 cm and a length of 8.7 cm (Figure \ref{fig:rock}(a)). For the fine resolution, a close-up view targeting the center of this area was scanned (Figure \ref{fig:rock}(b)). The high resolution was chosen to accurately capture the pore structure necessary for calculating the permeability through pore-scale flow simulation.

\section{Method}
\subsection{Overview of the upscaling workflow}
\begin{figure}
\includegraphics[width=\columnwidth]{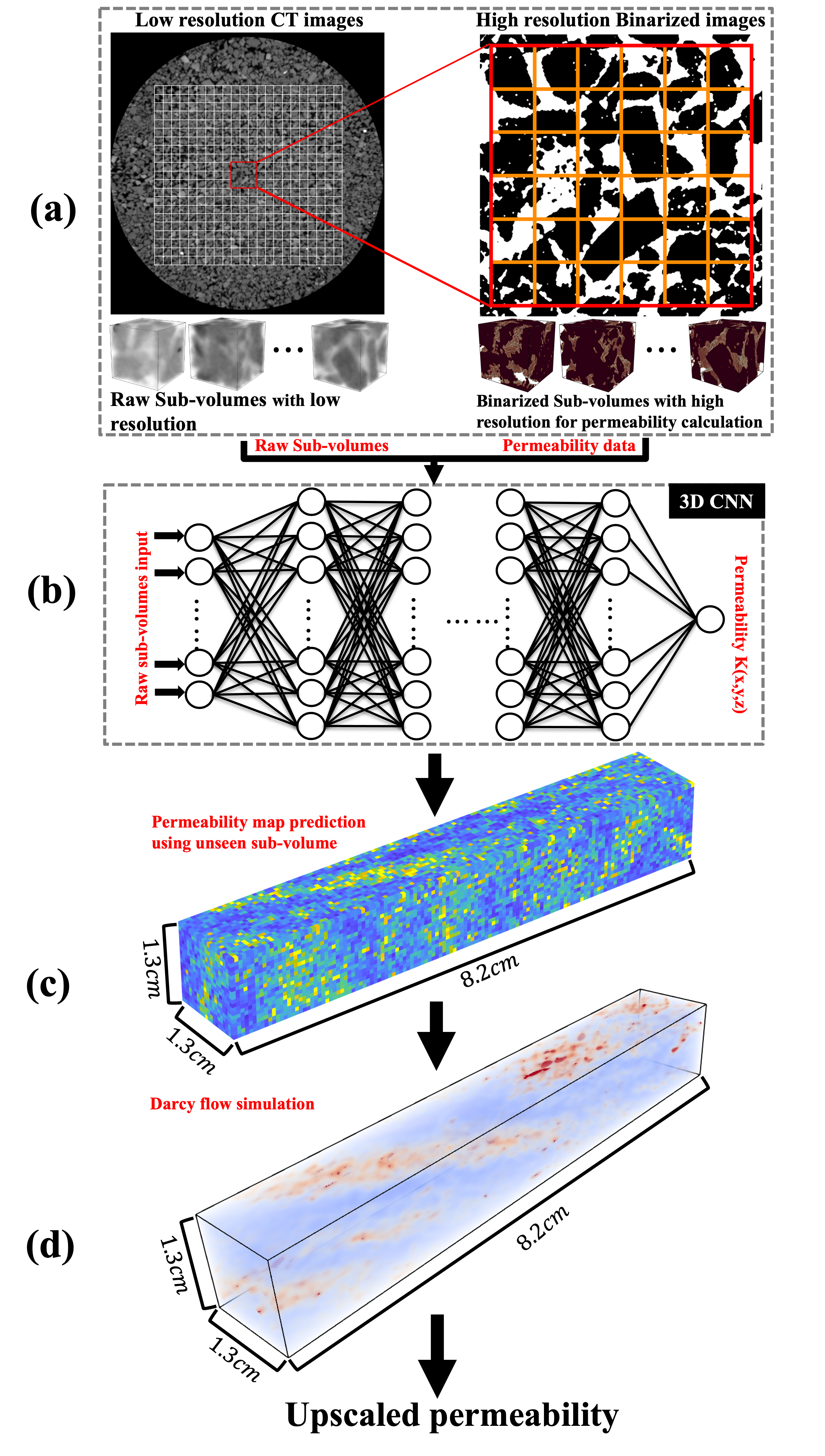}%
\caption{Permeability upscaling workflow: (a) CT scanning at different resolutions and creating training datasets; (b) Machine learning using raw CT data and their permeability datasets; (c) Predicting permeability maps of the entire core using the trained CNN; (d) Calculating flow flux distribution using the Darcy flow solver. The red box in (a) indicates the selected region for high-resolution analysis, and the orange grid lines show how that region is further divided into sub-volumes for permeability computation.}\label{fig:workflow}
\end{figure}

This study utilizes the same permeability upscaling approach as \citet{jiang2023upscaling}, which is divided into three main components: (a) Rock samples are imaged at both high and low resolutions. The binarized high-resolution CT images are fed into an LBM flow solver to obtain rock permeability data. Concurrently, the corresponding subvolumes are extracted from the low-resolution CT images. These subvolumes and permeability datasets are then employed to train a CNN. (b) The trained CNN is subsequently used to predict the permeability distribution throughout the entire rock sample. (c) Finally, the predicted permeability distribution is utilized to compute the upscaled permeability using a Darcy flow solver. 
Our workflow involves calculating permeability from high-resolution micro-CT images and aligning these data with corresponding larger-scale, low-resolution images. This approach leverages the precise permeability data from high-resolution images while benefiting from the broader field of view provided by low-resolution images, enabling the capture of spatial heterogeneity. Deep learning regression techniques enable the prediction and mapping of permeability variations in large-scale, low-resolution images directly from raw image data, which can be input into the CNN without prior binarization.
The detailed workflow includes the following steps:

\begin{enumerate}
\item Perform micro-CT coarse scanning of the entire core sample (8.7 cm in length) at a low resolution of 20$\mu m$/voxel (left-hand side image in Figure \ref{fig:workflow}(a));
\item Select the region for close-up imaging (20 mm wide) and conduct high-resolution micro-CT scanning at the pore scale (3$\mu$m/voxel), then reduce noise and segment the fine-scanned micro-CT images (right-hand side image in Figure \ref{fig:workflow}(a));
\item Extract subvolumes (grids in Figure \ref{fig:workflow}(a) right) from the segmented high-resolution images;
\item Calculate permeability properties for the extracted high-resolution subvolumes (dark red cubes in Figure \ref{fig:workflow}(a) right) using the LBM flow solver;
\item Register the overlapping subvolumes at the two different resolutions (Figure \ref{fig:workflow}(a));
\item Link the permeability data to the corresponding low-resolution subvolumes in the coarse-scanned 3D CT raw images (red squares in the center and gray-scale cubes of Figure \ref{fig:workflow}(a) left);
\item Create a training dataset composed of the extracted raw subvolumes and their corresponding permeability data;
\item Train the CNN model (Figure \ref{fig:workflow}(b)) using the prepared dataset;
\item Extract raw subvolumes from the outer region of the large-scale rock sample from the low-resolution CT images (white grids in Figure \ref{fig:workflow}(a) left) and predict the permeability for these subvolumes using the trained CNN (Figure \ref{fig:workflow}(c));
\item Calculate the flow flux within the whole core using the predicted permeability map by the MATLAB Reservoir Simulation Toolbox (MRST) \citep{lie2019introduction} Darcy flow solver (Figure \ref{fig:workflow}(d)); 
\item Estimate the upscaled permeability for the entire rock core.
\end{enumerate}

\subsection{Training Dataset Acquisition}
\label{secData}
Image processing of micro-CT images is essential for accurate pore-scale flow direct numerical simulation. It ensures the precise representation of pore geometry, reduces noise, and effectively segments pore and solid phases, all of which are critical for defining accurate computational domains. In this study, a non-local mean filter was applied to the high-resolution CT images to enhance image quality. This filter effectively reduces salt-and-pepper noise, resulting in smoother and more uniform images while preserving boundaries. Then, these high-resolution images underwent a segmentation process to extract the exact pore geometry. The segmentation was performed using a simple threshold algorithm based on the grayscale histogram of the image. The threshold values were selected considering the local minimum values of the histogram. It is important to note that only high-resolution images require processing and segmentation for direct flow simulation. The original low-resolution raw images can be directly utilized for deep learning. Finally, a binarized pore-geometry model with voxel dimensions of $660\times650\times17230$ voxels was reconstructed from the cylindrical sample area (Figure \ref{fig:rock}(b)). 

\begin{figure}[t]
\includegraphics[width=\columnwidth]{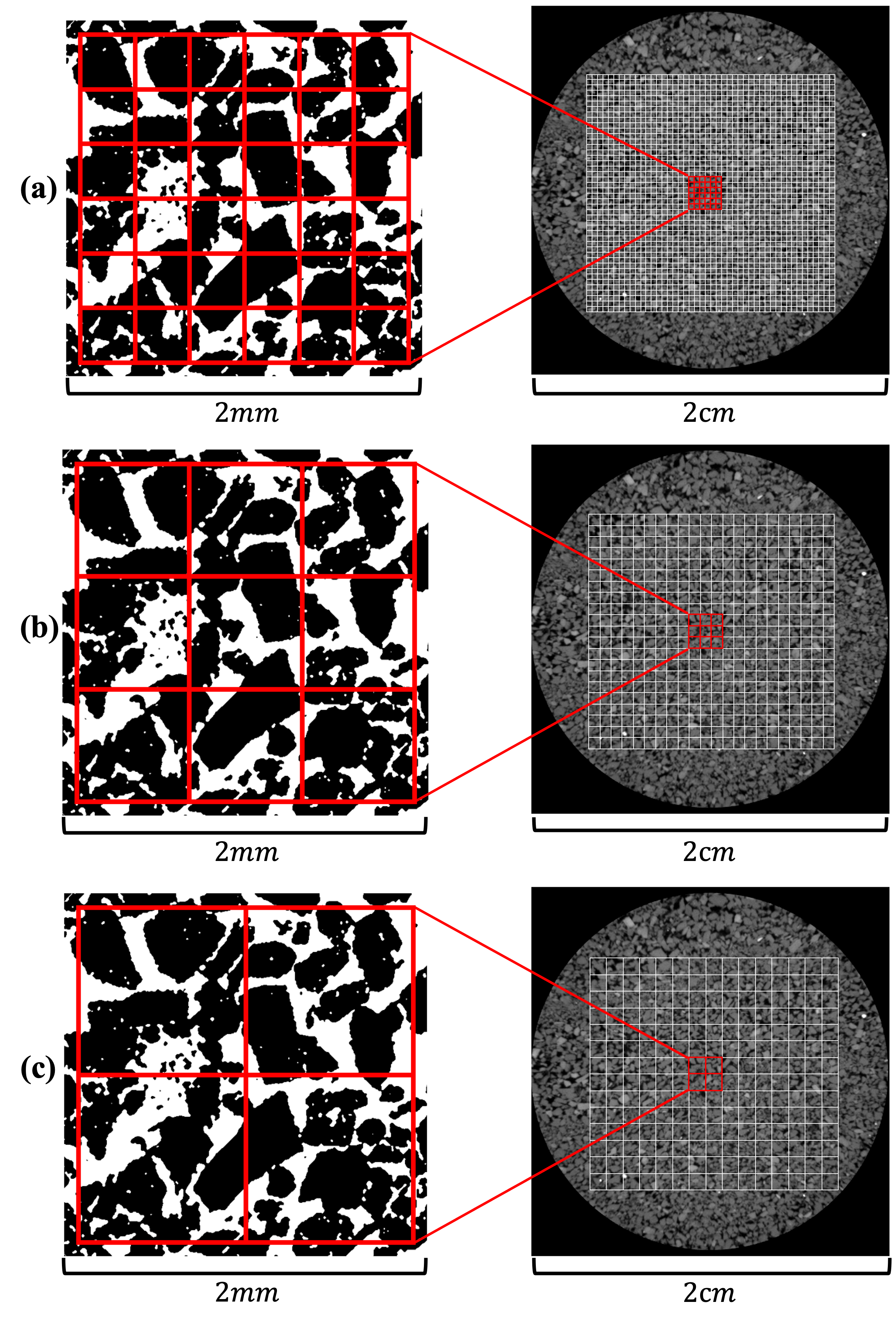}
\caption{Illustration of the subvolume extraction process for the binarized high-resolution CT images (left) and low-resolution CT images (right) with three different sizes: (a) size $100^3$(Case I), (b) size $200^3$(Case II), and (c) size $300^3$(Case III)}.
\label{fig:CT}
\end{figure}

The binarized model was subsequently partitioned into subvolumes of $100^3$ (Case I), $200^3$ (Case II), and $300^3$ (Case III) voxels (left-hand side in Figure \ref{fig:CT}(a),(b),(c)). To meet the requirements for large datasets, subvolumes were allowed to overlap by 50 voxels during the extraction process. 
The permeability of these subvolumes was calculated by the LBM flow solver by simulating single-phase flow in the z-direction. Therefore, the calculated permeability is a scalar value representing the effective permeability along the z-axis. Porosity was computed arithmetically from the binarized high-resolution CT images.
Corresponding low-resolution subvolumes at the same locations in the sample were then extracted directly from the original coarse-scanned images. Given the 6.7-fold resolution difference between the high and low-resolution scans, the dimensions of the low-resolution subvolumes were $15^3$, $31^3$, and $46^3$ voxels (red square in the center of right-hand side in Figure \ref{fig:CT}(a),(b),(c)). Thus, we acquired a dataset for deep learning comprising pairs of low-resolution subvolumes and their corresponding permeability or porosity values. The total number of subvolumes for each size is summarized in Table \ref{table:1}. In our study, 90\% of the data were used for training, while the remaining 10\% were used for testing.

To further characterize the representativeness of the extracted subvolumes, we also examined their average pore counts using pore network modeling (PNM) analysis (Table~\ref{table:1}). The results show that subvolumes of $100^3$, $200^3$, and $300^3$ voxels contained on average approximately 74, 421, and 1240 pore bodies, respectively. As expected, larger subvolumes encompass a greater number of pores. 

Since voxel-based dimensions (e.g., $100^3$, $200^3$, $300^3$) are inherently dependent on CT image resolution and rock grain size, they lack general applicability across different rock types and imaging conditions. For example, a $100^3$ voxel subvolume at 3$\mu$m resolution contains about 74 pores in Boise sandstone, whereas rocks with coarser grains would require larger voxel sizes to capture a comparable number of pores. Therefore, the average pore count provides a more robust and dimensionless indicator of subvolume representativeness, independent of voxel resolution or rock type. This metric will be used in subsequent discussions as a general indicator of subvolume representativeness.

\begin{figure}[t]
\includegraphics[width=\columnwidth]{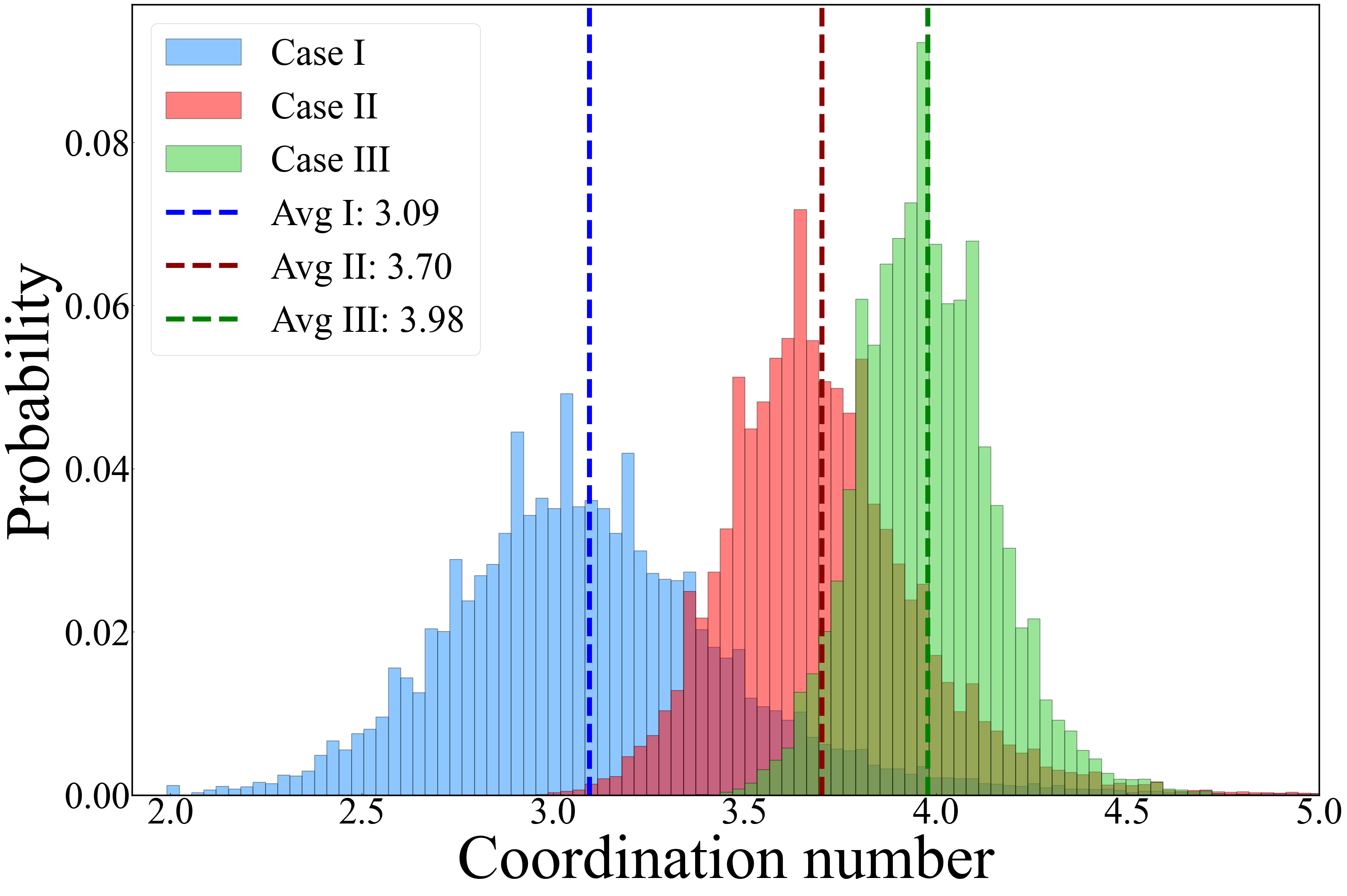}%
\caption{Coordination number distributions for the pore networks extracted from each subvolume size. A higher coordination number indicates better connectivity and flow continuity in the network. Case III shows the highest average value and narrowest distribution.}
\label{fig:CoordNumber}
\end{figure}

To further investigate the structural representativity of each subvolume size, we analyzed the coordination number distributions derived from the pore network models (Figure~\ref{fig:CoordNumber}). 
The coordination number reflects the average number of connections (throats) per pore body, which is a key topological property affecting flow connectivity. 
 Case~III exhibits the highest average coordination number (3.98), with a sharp peak and narrow distribution, indicating a well-connected and topologically stable network. 
In contrast, Case~I shows a broader distribution centered around a lower average (3.09), suggesting more local heterogeneity and disconnected features.
These results imply that larger subvolumes, such as Case~III, provide not only statistical representativity in terms of pore size and porosity but also more stable flow pathways at the network level.

\begin{table}[ht]
\centering
\begin{tabular}{c c c c} 
\hline
   Case & I & II & III \\
 \hline
 \begin{tabular}{c} 
 Fine subvolumes size \\
 (Voxel, \(20 \mu \mathrm{m} /\)Voxel)
 \end{tabular} & \(100^3\) & \(200^3\) & \(300^3\) \\
 \hline
 \begin{tabular}{c} 
 Coarse subvolumes size \\
 (Voxel, \(3 \mu \mathrm{m} /\)Voxel)
 \end{tabular} & \(15^3\) & \(31^3\) & \(46^3\) \\
 \hline
 Number of training data & 48960 & 34000 & 21120 \\
 \hline
 \begin{tabular}{c}
 Averaged number of pores  \\
 (based on PNM analysis)
 \end{tabular} & 74 & 421 & 1240 \\
 \hline
\end{tabular}
\caption{Subvolume's dimensions, training data number, and average pore count for the high and low-resolution models in three cases.}
\label{table:1}
\end{table}

\section{Results and discussion}




\subsection{Permeability results of the high-resolution region by LBM}
\label{subsecPer}
\begin{figure}[t]
\includegraphics[width=\columnwidth]{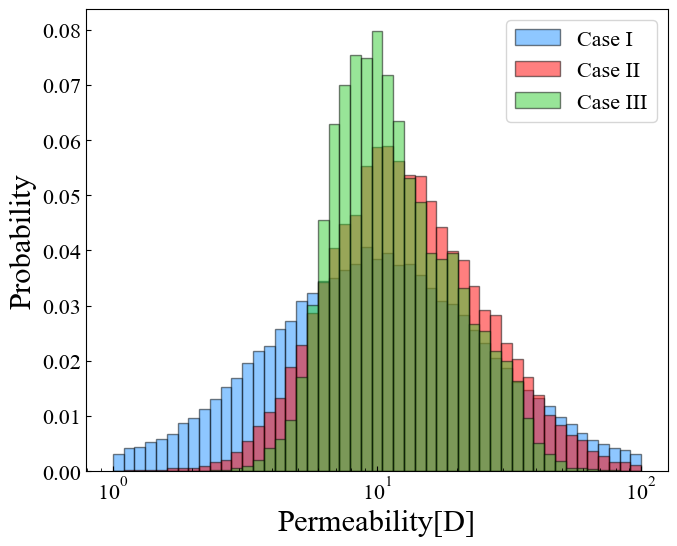}%
\caption{Histograms of the calculated permeability in logarithmic scale for Case I, II, and III}.
\label{fig:histgramC}
\end{figure}

The permeability distributions for Case I, II, and III are presented in Figure \ref{fig:histgramC}. The standard deviations of permeability for these subvolumes are 20.73, 12.88, and 8.05 Darcy, respectively. These findings indicate that the permeability distribution changes with subvolume size. Notably, larger subvolumes exhibit more concentrated permeability distributions, as reflected by their smaller standard deviations.

From a machine learning perspective, data diversity is a critical factor in achieving successful predictions. The histogram for Case I showed that permeability data are broadly distributed, indicating significant diversity. This increased diversity arises because smaller subvolumes highlight local heterogeneities in the rock, generating more varied data. At first glance, this may suggest that Case I subvolumes are most appropriate for accurate predictions. However, subvolumes of size 100 may include flow paths that do not connect within the larger sample or consist solely of minerals with minimal flow paths. These outlier subvolumes can lead to  overestimation or underestimation of permeability, resulting in inaccurate predictions.  
In contrast, as the subvolume size increases, data tend to converge within a narrower range (Figure~\ref{fig:histgramC}). This convergence occurs because larger subvolumes average out local heterogeneities in the rock, thereby reducing data diversity. Although data diversity diminishes with larger subvolumes, the chances of encountering outlier subvolumes that are completely composed of flow paths or entirely made up of minerals also reduce, resulting in more accurate permeability predictions. A reduction in the number of outlier subvolumes enhances the reliability of large-scale permeability map predictions, which also contributes to more accurate calculations of upscaled permeability in the Darcy flow solver.
Therefore, the trade-off between data diversity and permeability map prediction accuracy in machine learning models requires careful consideration. 
Choosing an appropriate subvolume size is essential for optimizing the balance between prediction accuracy and quality of training data, leading to more efficient upscaling of permeability.

\subsection{Accuracy of CNN for different subvolume sizes}
In this study, we used ResNet34, a type of CNN architecture, for permeability prediction based on its excellent performance \citep{jiang2023upscaling}. The training datasets mentioned in Section \ref{secData} were used to train the ResNet34 model. The model was trained for 300 epochs, achieving the convergence of the loss function. The parameters of the network that exhibited the best performance during training were stored for use in predictions.

\begin{figure}[t]
\includegraphics[width=\columnwidth]{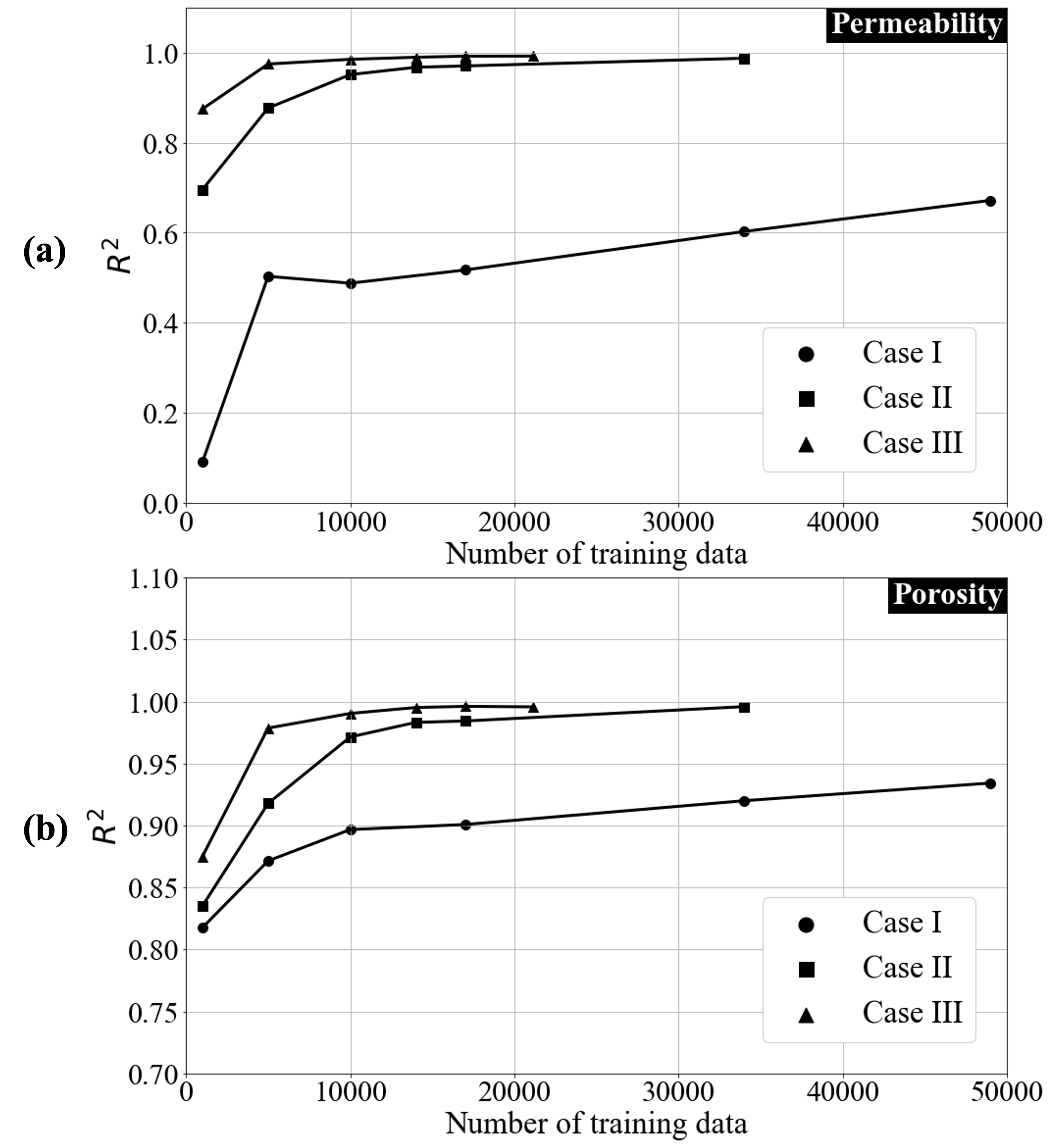}%
\caption{Relationship between the number of training data and the learning accuracy $R^2$ in permeability (a) and porosity (b) prediction for Case I, II, and III.}
\label{fig:Permeability_Porosity_R2}
\end{figure}

First, we investigated the effect of training data size on prediction accuracy. We used the coefficient of determination (\(R^2\)) to evaluate prediction accuracy, which is a standard metric in regression analysis \citep{draper1998applied}.  An \(R^2\) value near 1 indicates a good model fit, while a value close to 0 indicates a poor fit. Figure \ref{fig:Permeability_Porosity_R2} showed that the \(R^2\) value increases as the number of training data points increases for all three subvolume sizes, indicating improved learning accuracy. The prediction results of permeability (Figure \ref{fig:Permeability_Porosity_R2}(a)) showed that Case III achieves the highest  $R^2$  value, close to 1.0, with fewer training data, and it stabilizes quickly as the number of training data increases. Similarly, Case II reaches a high  $R^2$  value but requires slightly more data to approach the maximum accuracy compared to Case III. In contrast, Case I initially displays a much lower  $R^2$  value. As more training data is added, the performance steadily improves. However, even with 48,960 data points, Case I does not reach the accuracy levels observed with the larger subvolumes.
In Case I, the prediction accuracy does not fully converge, while for case II, convergence is achieved with approximately 20,000 data points. For Case III, it reaches satisfactory prediction accuracy with as few as 10,000 data points.
Similarly, for porosity predictions (Figure \ref{fig:Permeability_Porosity_R2}(b)), Case III achieves near-perfect accuracy with a relatively small amount of data, closely followed by Case II.  The smaller subvolume, Case I, lags behind again, requiring more training data to reach comparable accuracy levels. The convergence rate for porosity prediction showed a similar trend to that of permeability prediction across different subvolume sizes. 
These results suggest that increasing the size of the subvolumes enhances the model’s ability to predict permeability and porosity accurately with fewer training data points. This may be due to the larger subvolumes providing more representative features of the underlying physical properties, leading to a more robust model.
While the use of larger subvolumes improves prediction accuracy and model robustness, it also leads to increased computational cost during model training. 
As shown in Table~\ref{table:training_time}, the training time for the ResNet34 model increases substantially with subvolume size: approximately 8.3 hours for Case~I ($100^3$), 39.3 hours for Case~II ($200^3$), and 89.8 hours for Case~III ($300^3$), using a single NVIDIA A6000 GPU. 
Additionally, memory consumption during training also increases with subvolume size due to the higher input dimensionality and feature complexity, though specific values may vary depending on implementation and batch size.

\begin{table}[!ht]
\centering
\begin{tabular}{ccccc}
\hline
Case & I & II & III \\
\hline
Subvolume size (voxel) &  $100^3$  &  $200^3$  &  $300^3$  \\
Training time (hr.) &  8.3  &  39.3  &  89.8  \\
Memory usage (GB) &  23.3  &  25.8  &  33.2 \\
\hline
\end{tabular}
\caption{CNN training time and memory usage for each case.}
\label{table:training_time}
\end{table}

\begin{figure}[t]
\includegraphics[width=\columnwidth]{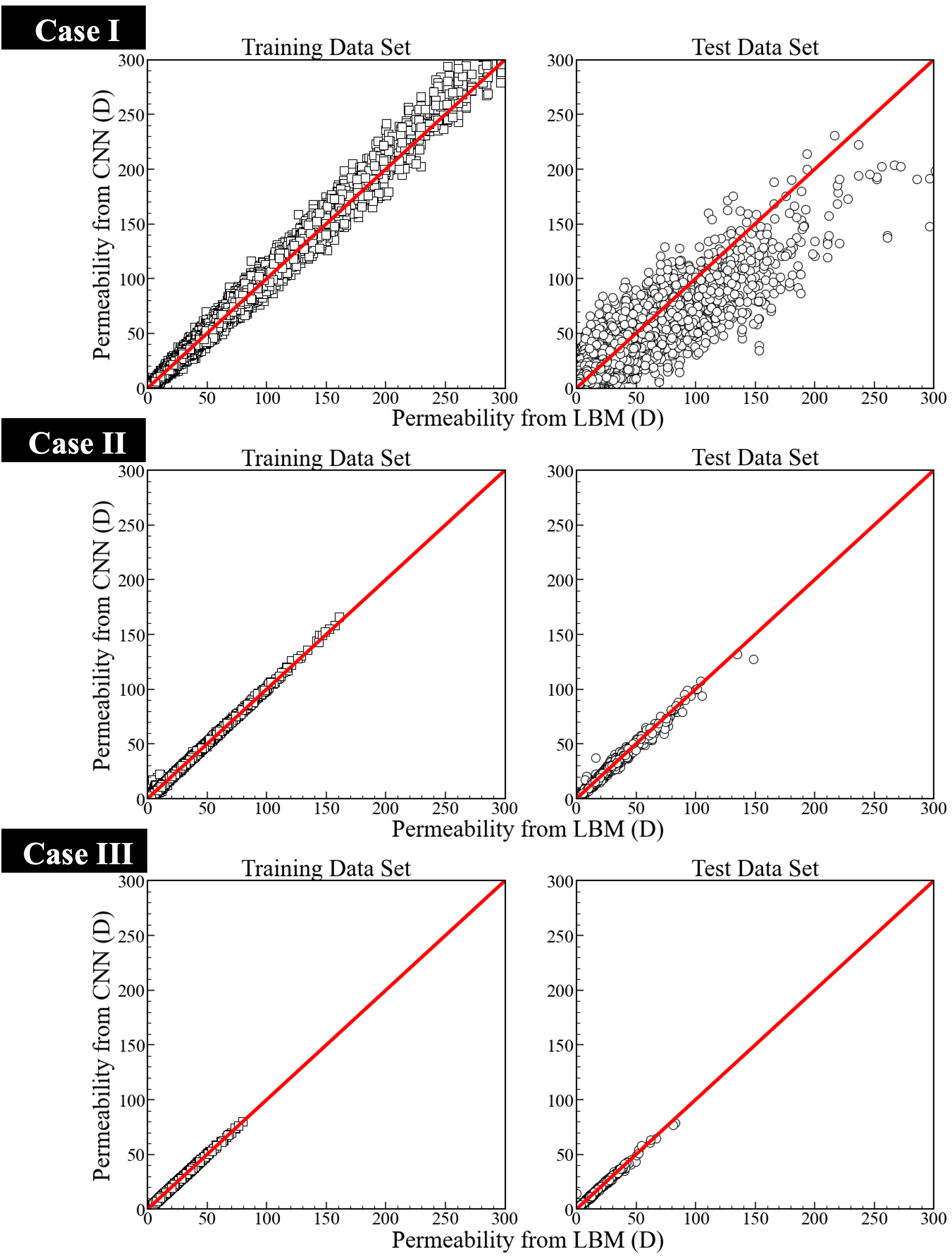}
\caption{Permeability values from LBM are plotted against the permeability from CNN for Case I, II, and III. The red diagonal line signifies the perfect match between the predicted permeability and the actual permeability derived from simulations.}
\label{fig:AI-Permeability-plot}
\end{figure}

\begin{figure}[t]
\includegraphics[width=\columnwidth]{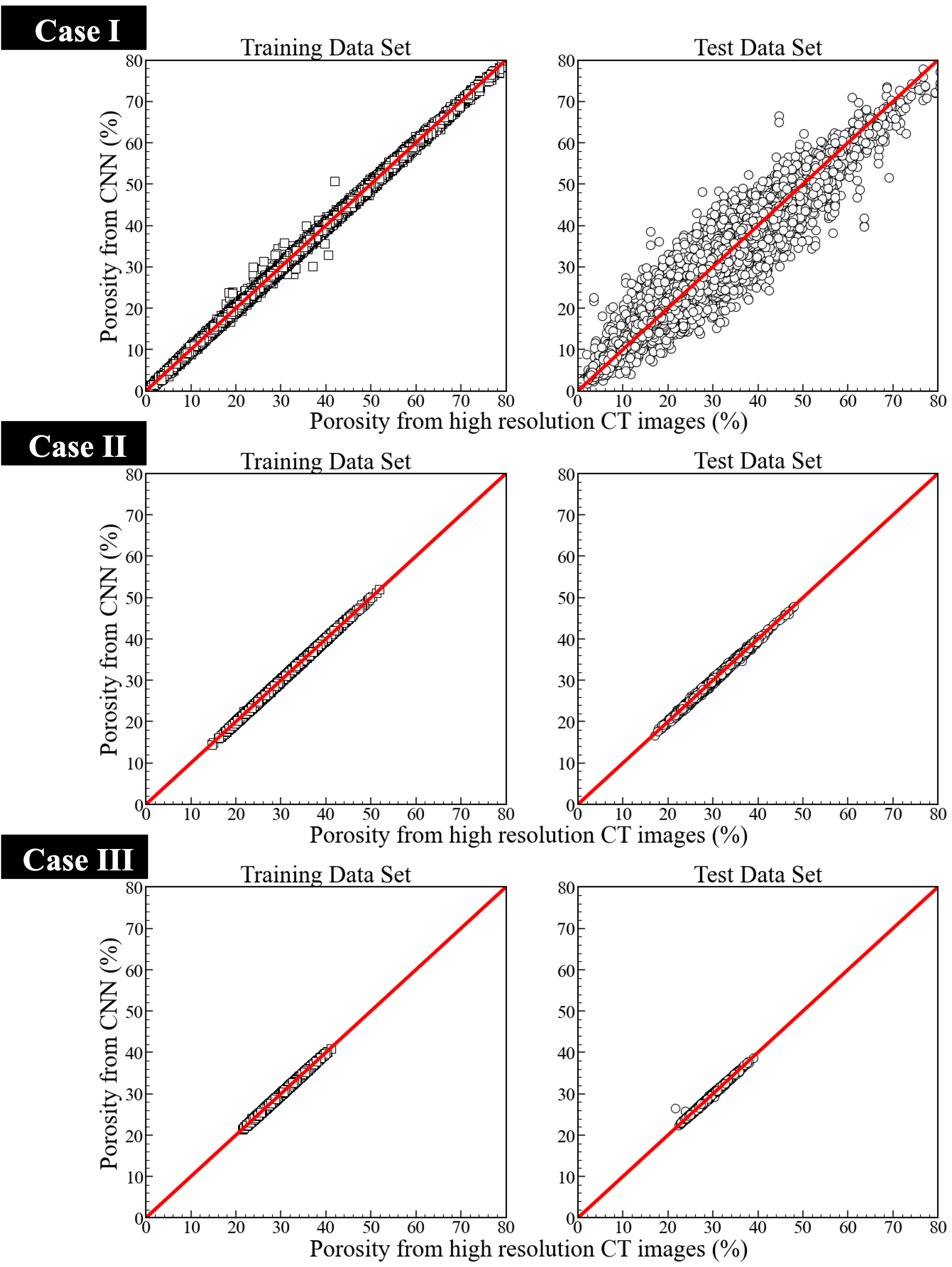}
\caption{Porosity values from high resolution CT images are plotted against the porosity from CNN for Case I, II, and III. The red diagonal line signifies the perfect match between the predicted porosity and the actual porosity derived from subvolumes.}
\label{fig:AI-Porosity-plot}
\end{figure}

Subsequently, we evaluated the performances of prediction at the point of convergence for three different subvolume size scenarios. The predicted values were plotted against the actual permeability obtained from the LBM direct flow simulation and the actual porosity obtained from the same segmented subvolume models (Figure \ref{fig:AI-Permeability-plot} and \ref{fig:AI-Porosity-plot}).
For Case I, there is a significant prediction error and a large variation in the predicted permeability. 
This is because smaller subvolumes contain fewer features of the rock, and the characteristics between each subdomain can vary greatly. 
In contrast, for Case II, the predictions are more tightly clustered around the diagonal line in both training and test datasets, suggesting improved accuracy and better generalization compared to Case I. The most accurate predictions are seen with Case III, where the permeability values from both the training and test datasets almost perfectly align with the simulated values,
making Case III the most reliable among the three cases for permeability prediction. This improvement is due to the larger subvolumes capturing more features of the rock and covering a broader range, leading to less variation between subdomains. 
Regarding porosity prediction, Case I also exhibits a fairly large prediction error (Figure \ref{fig:AI-Porosity-plot}). For Case II and III, the prediction performance improves, with Case III showing predictions very close to the actual porosity. The accuracy of porosity predictions is notably higher than that of permeability predictions. This difference arises because porosity is primarily determined by the gray-scale histogram, whereas permeability is also influenced by the connectivity of the pore space, such as the tortuosity of the flow paths. Consequently, porosity is inherently simpler to characterize using deep learning techniques.

These findings highlight the significant influence of both training data size and subvolume dimensions on the accuracy of permeability and porosity predictions. In particular, larger subvolumes tended to yield better performance by providing more representative training data. These observations suggest that subvolume size is an important factor to consider in constructing effective training datasets for upscaling applications.

\subsection{Permeability map predicted by the trained CNN}
\begin{figure}[t]
\includegraphics[width=\columnwidth]{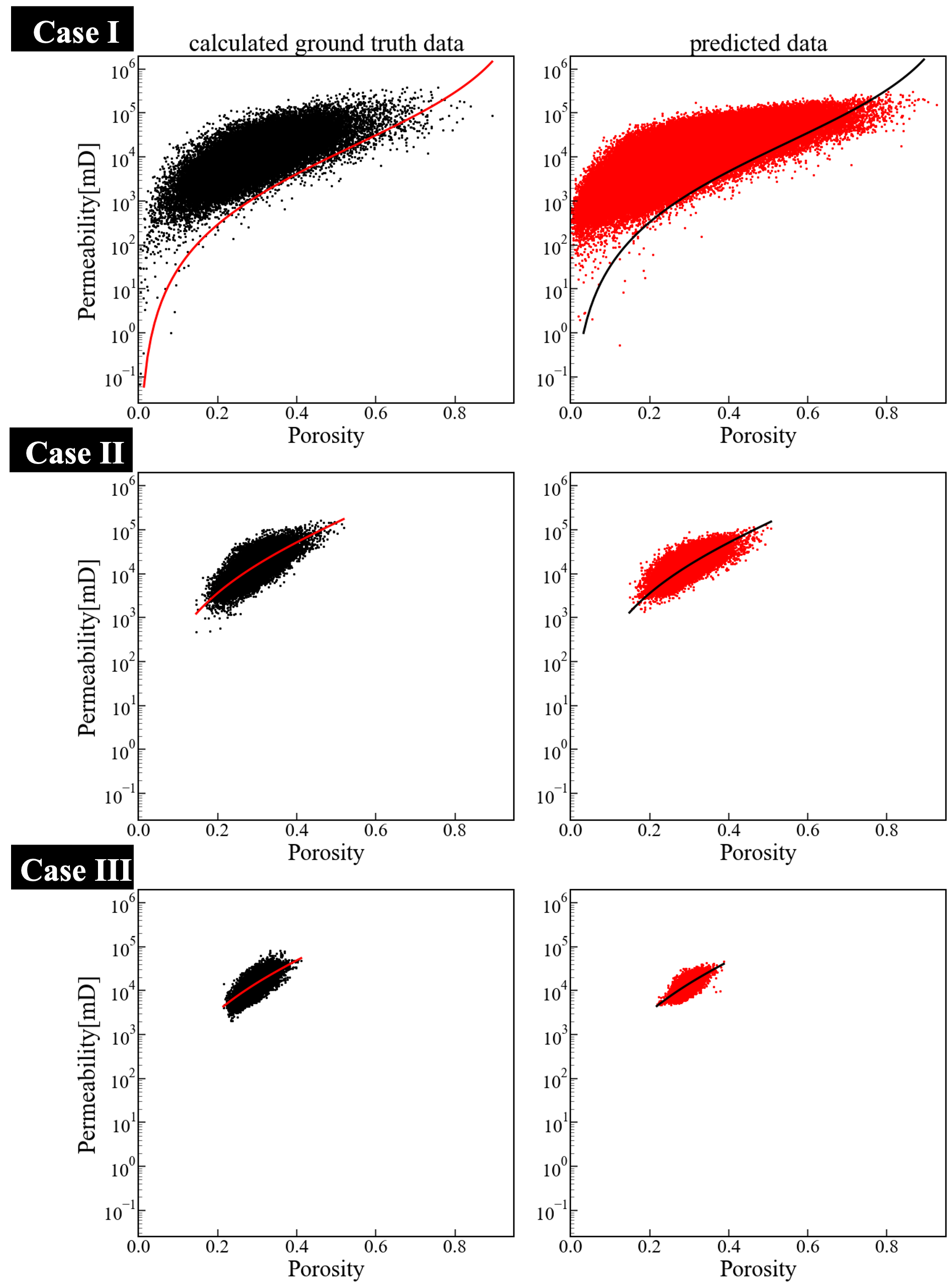}
\caption{Relationship between the permeability and the porosity for Case I, II, and III. The solid line in the figures shows the results predicted by the KC model.} 
\label{fig:KC}
\end{figure}

We then used the trained CNN to predict the permeability and porosity of subvolumes in the unseen peripheral parts of the rock core (white grid part in Figure \ref{fig:workflow}(a) left).
Since permeability values are primarily influenced by porosity, we examined the permeability-porosity relationship for the predicted results. The scatter plots of predicted permeability versus porosity showed patterns similar to those of the calculated data for each size (Figure \ref{fig:KC}), demonstrating the reliability of our trained neural network.
The most basic empirical model to describe the relationship between permeability and porosity is the Kozeny-Carman (KC) model \citep{carman1937fluid}. The classical KC model \citep{costa2006permeability} can be expressed as :
\begin{equation}
    K=C \frac{\phi^{3}}{(1-\phi)^2}
\end{equation}
where, $C$ is the parameter related to the geometrical properties and $\phi$ denotes the porosity. 
We employ this KC model to fit our results. In Figure \ref{fig:KC}, the fitting curve of the KC model for Case I exhibits a downward shift. 
This deviation arises from the presence of numerous outliers within the 100-sized subvolumes, leading to a bias in the fitting outcomes.
As the subvolume size increases to 200 and 300 voxels, both the ground truth and predicted data exhibit tighter clustering around the KC model line, with the predicted data closely mirroring the ground truth. This suggests that the larger subvolume size enhances the model’s ability to accurately capture the relationship between permeability and porosity, leading to more reliable predictions with less variability compared to the smaller subvolume size.
Our data-driven approach outperforms the empirical KC model, which only correlates permeability with porosity, neglecting the influence of pore microstructure and flow path connectivity. The simple KC model inadequately capture the complexity of the permeability-porosity relationship, especially in heterogeneous samples. An essential advantage of our approach is leveraging machine learning to extract detailed information, enabling a more comprehensive reproduction of the permeability-porosity relationship distribution considering heterogeneity. 
The permeability map predicted from low-resolution 3D images by the neural network serves as a basis for further permeability upscaling.


\begin{figure}[t]
\includegraphics[width=\columnwidth]{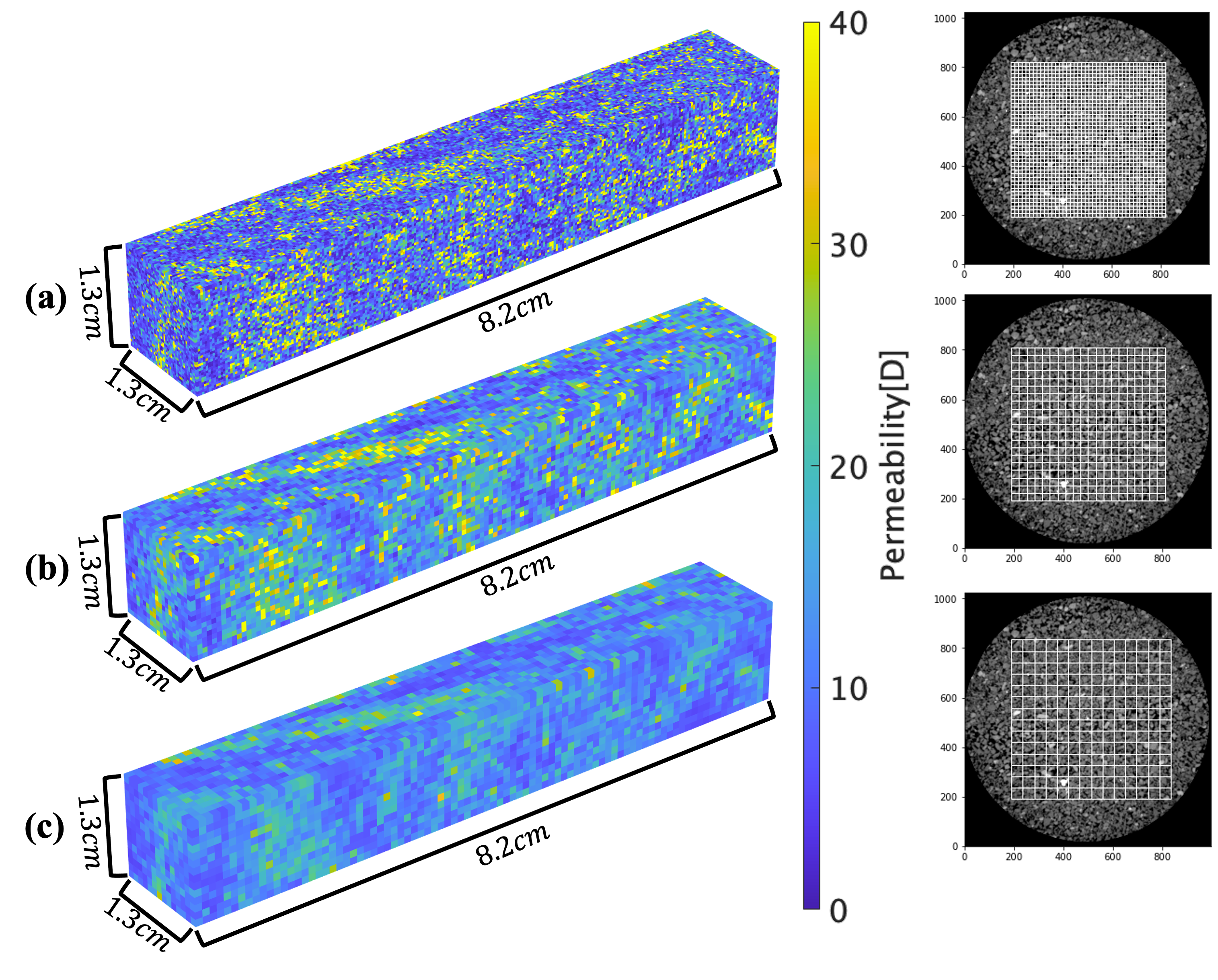}
\caption{Permeability maps for the whole cores (left figures) and the cut rectangular region from the original cylindrical sample for the Darcy flow calculation (white boxes in the right figures): (a) Case I, (b) Case II, (c) Case III.} 
\label{fig:PermeabilityMap}
\end{figure}

The predicted permeability map (axial direction of the core) that illustrates the spatial distribution of permeability across the core samples at different subvolume resolutions is shown in Figure \ref{fig:PermeabilityMap}. 
From Case~I (Figure~\ref{fig:PermeabilityMap}(a)) to Case~III (Figure~\ref{fig:PermeabilityMap}(c)), the maps reveal progressively clearer macroscopic patterns of permeability distribution. 
In Case~I, the map appears highly granular with significant variability at the small scale. 
As the subvolume size increases, this small-scale variability decreases, larger-scale flow structures become more visible. 
Case III exhibits the smoothest distribution with a more uniform and coherent structure. This indicates that larger subvolumes may help to better capture the macroscopic flow characteristics by minimizing the influence of small-scale heterogeneities that might otherwise act as noise.

\begin{figure}[t]
\includegraphics[width=\columnwidth]{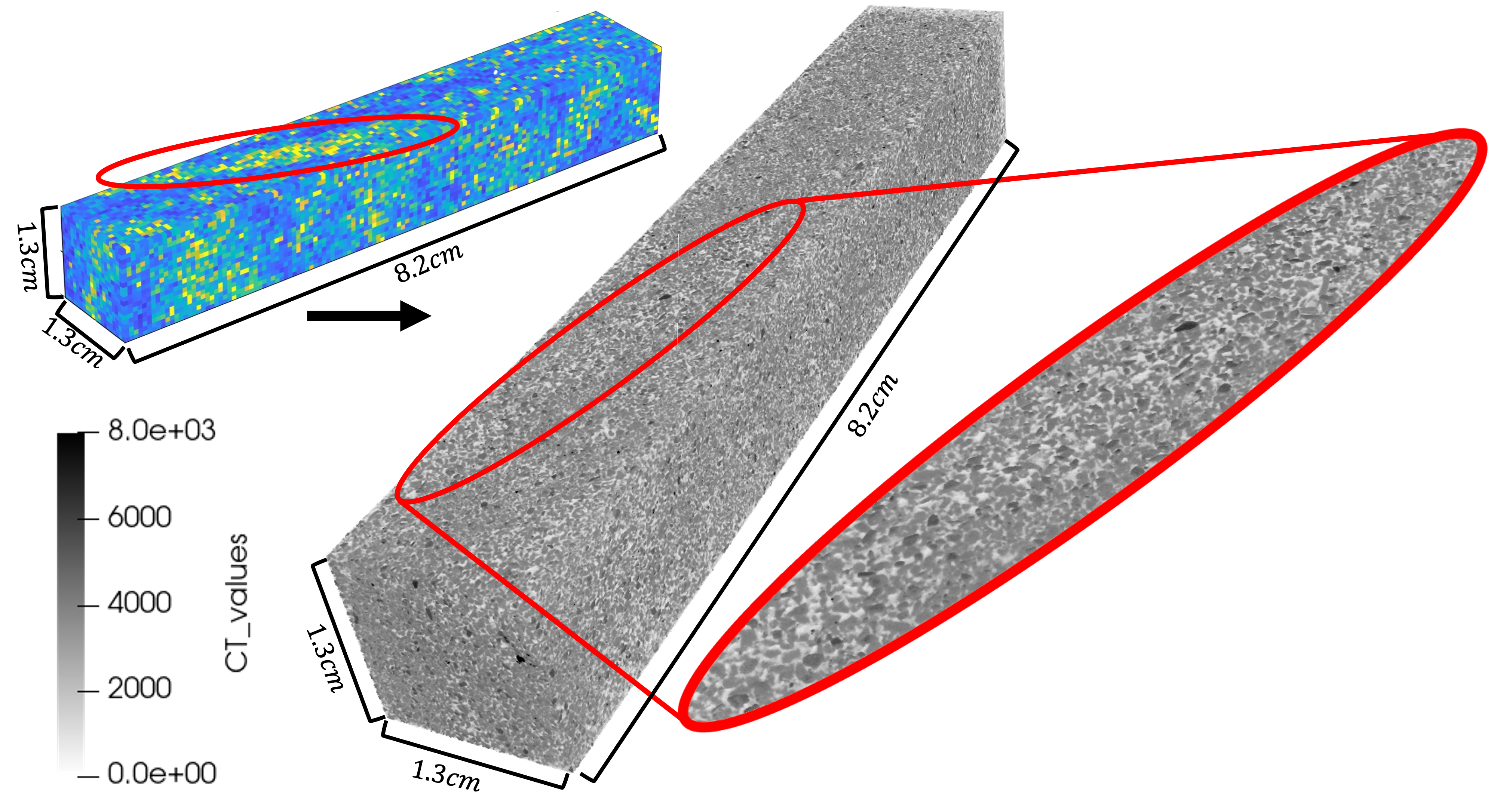}
\caption{Permeability map and the corresponding raw CT rock model.} 
\label{fig:raw_model}
\end{figure}

Figure \ref{fig:raw_model} presents a comparison between a permeability map and the corresponding raw CT rock model, which demonstrates the spatial alignment between the permeability map and the underlying rock structure as captured by the CT scan. When comparing these regions to the corresponding areas in the CT rock model, there is a clear correlation between the permeability variations and the rock’s microstructure. The region highlighted in red demonstrates that zones with less dense rock structures in the CT scan match areas of higher permeability in the map. Conversely, denser zones in the CT scan correspond to lower permeability regions in the map. This agreement supports the accuracy of our permeability map, confirming the effectiveness of our method.

\subsection{Upscaled permeability of the whole core}
\begin{figure}[t]
\includegraphics[width=\columnwidth]{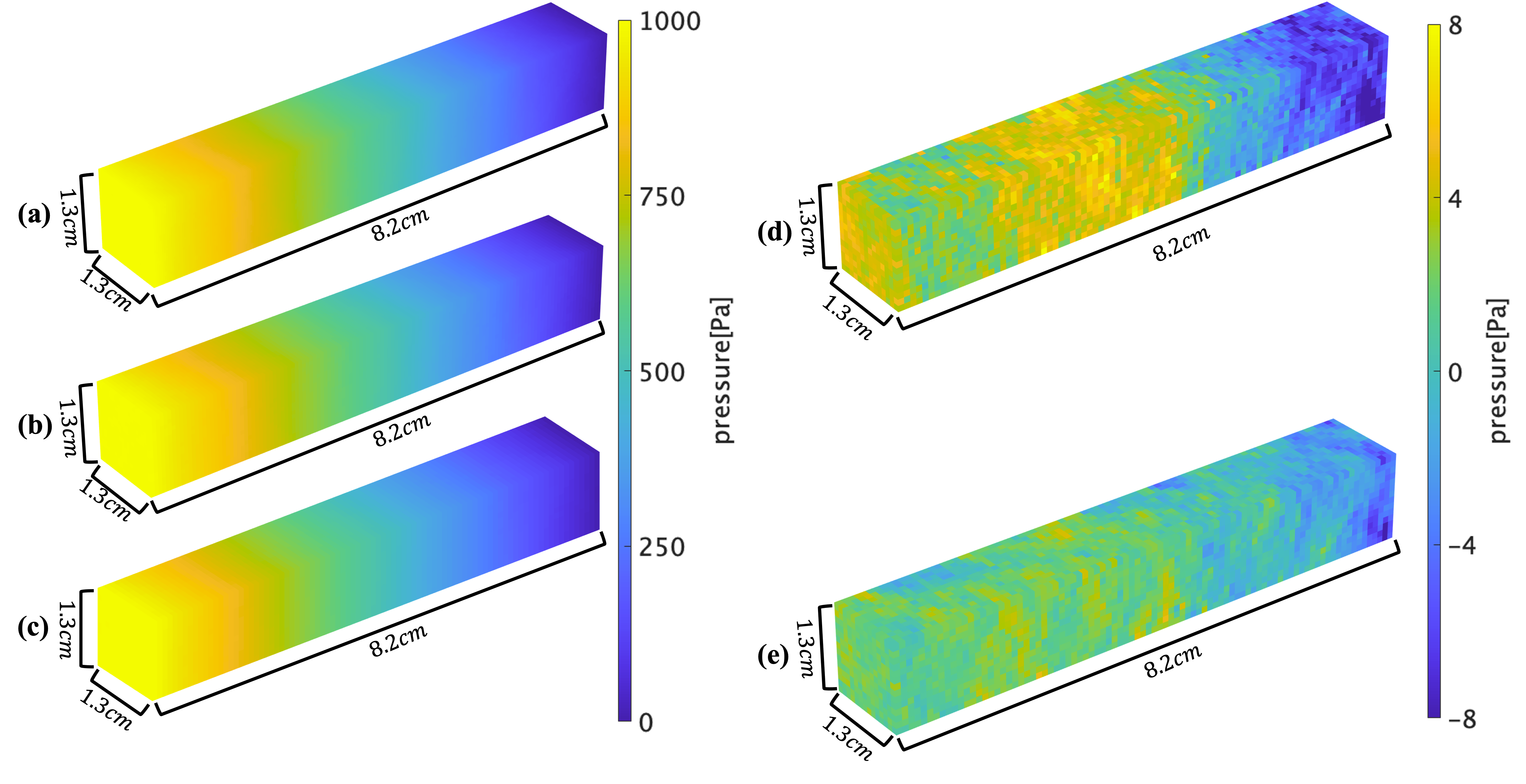}
\caption{Pressure distributions of the whole core obtained from the Darcy flow solver: (a) Case I, (b) Case II, (c) Case III, (d) pressure difference between Case I and III, (e) pressure difference between Case II and III.} 
\label{fig:PressureMap}
\end{figure}

To consider the anisotropy of permeability, the trained CNN was applied to the subvolumes rotated along the x, y, and z axes. The estimated permeabilities in each principal direction ($K_{xx}$, $K_{yy}$, $K_{zz}$) correspond to the diagonal elements of the permeability tensor. The off-diagonal components were neglected as their contribution to the overall permeability is minimal under axis-aligned flow conditions and has negligible impact on upscaled flow behavior.
Then the obtained permeability maps were used to calculate the pressure distribution of the entire sample using the Darcy flow solver. As the computational domain for the Darcy flow solver, rectangular cuboid regions with grid sizes of 42×42×267, 20×20×130, and 14×14×88 were extracted from the same region of the core sample for Case I, II, and III, respectively (white square of right-hand side image in Figure \ref{fig:CT}(a),(b),(c)). Each grid cell was assigned the predicted permeability values in three directions. Fixed pressure boundary conditions were imposed on the inlet and outlet sides to drive the fluid, while no-flow conditions were applied to the rest of boundaries.
Finally, the flow velocities and pressures within the entire core were obtained by solving the linear system constructed using the two-point flux approximation method \citep{lie2019introduction}.
The pressure distributions for all three cases showed a general trend of decreasing pressure from one end of the core to the other (Figure \ref{fig:PressureMap}(a), (b) and (c)).

To provide a detailed comparison of the pressure distribution results for each case, we computed the pressure differences by subtracting the pressure distribution of Case III from that of Case I (Figure \ref{fig:PressureMap}(d)) and from that of Case II (Figure \ref{fig:PressureMap}(e)).
The pressure difference maps reveal that the pressures for Case I and II are generally higher from the inlet to the midsection of the core flow direction compared to Case III, whereas the pressure near the outlet tends to be lower than that of Case III. The results also indicate that Case I, with the smallest subvolume, exhibits more local pressure fluctuations because of the large local permeability variations due to the influence of small-scale heterogeneity.
For Case II, while the pressure distribution becomes more uniform, some local variations remain when compared to Case III. In contrast, Case III exhibits minimal pressure variability, leading to a smoother overall distribution. Enlarging the subvolume size helps to average out the influence of small-scale heterogeneity, resulting in more uniform flow resistance. This indicates that increasing subvolume size improves the overall uniformity of the pressure distribution.

\begin{figure}[t]
\includegraphics[width=\columnwidth]{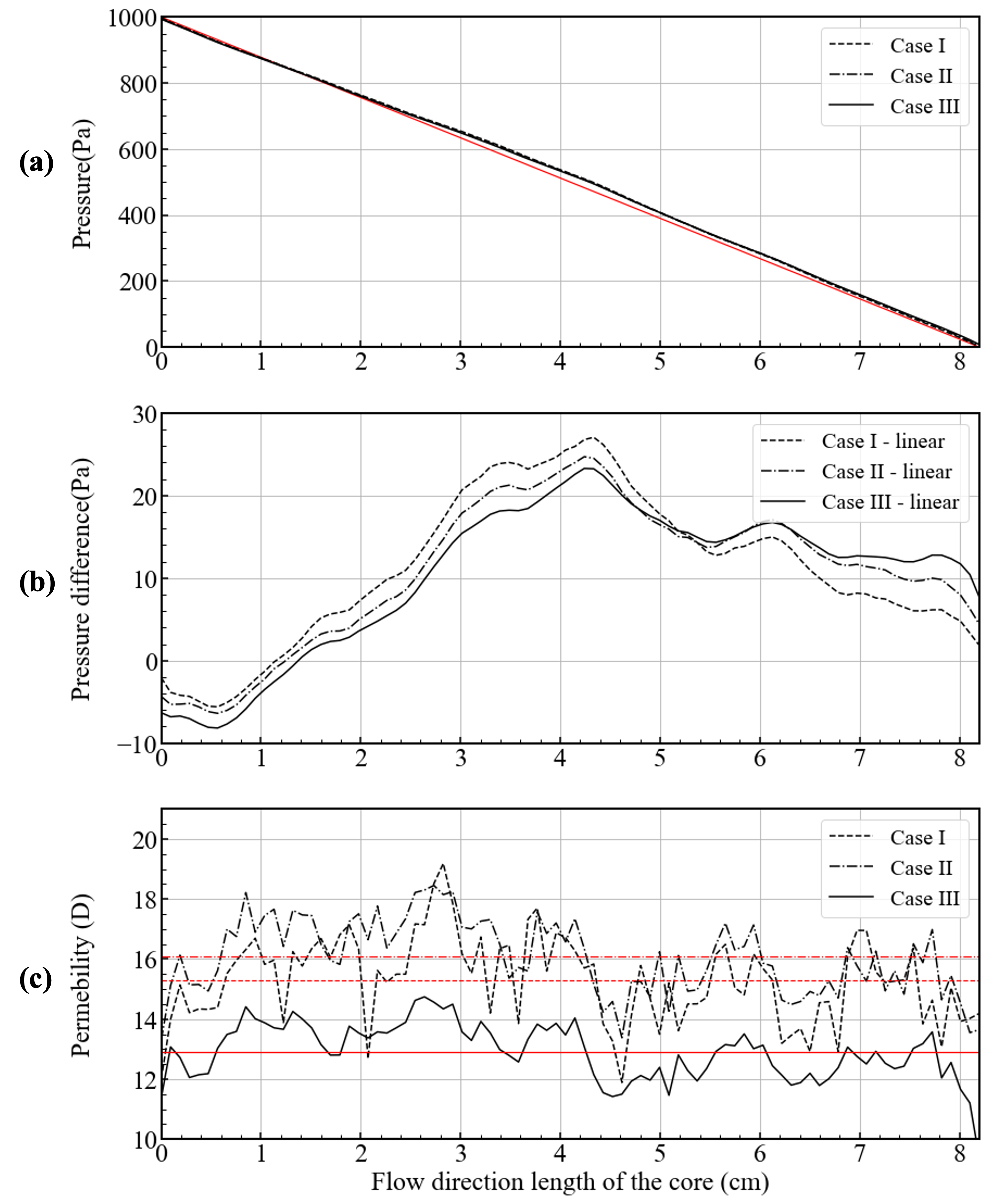}
\caption{The variation of properties along the flow direction of the core: (a)pressure (the red line in the figure represents the ideal linear pressure distribution), (b)pressure difference between averaged pressure and ideal linear pressure, (c) permeability (the red line in the figure represents the average permeability)} 
\label{fig:P_K_P_xplot}
\end{figure}

Under the assumption of a perfectly homogeneous porous medium, the pressure is expected to decrease linearly from the inlet to the outlet (Figure \ref{fig:P_K_P_xplot}(a), red line). To quantitatively evaluate the impact of heterogeneity for varying subvolume sizes, we calculated the difference between the simulated average pressure and the theoretically linear pressure distribution at each cross-section along the flow direction of the core (Figure \ref{fig:P_K_P_xplot}(b)).
The pressure difference exhibits a distinct pattern: it decreases in the inlet region, increases between 0.5 cm and 4.4 cm, and then declines again toward the outlet. This behavior strongly correlates with variations in permeability. In the region below 0.5 cm, permeability is lower than the average value (Figure \ref{fig:P_K_P_xplot}(c), red line), which increases resistance to fluid flow. As a result, a greater pressure gradient is required to drive the fluid, leading to a steeper decline in the averaged pressure compared to the ideal linear case. 
Between 0.5 cm and 4.4 cm, permeability exceeds the average value, indicating relatively high permeability that facilitates fluid movement. This reduces the required pressure gradient, causing a more gradual decrease of the averaged pressure compared to the ideal linear case.
At approximately 4.4 cm, a sharp reduction in permeability occurs, and from this point to the outlet, permeability remains below the average value. This increased resistance to flow necessitates a larger pressure gradient, causing the averaged pressure to drop more steeply relative to the ideal linear case.

\begin{figure}[t]
\includegraphics[width=\columnwidth]{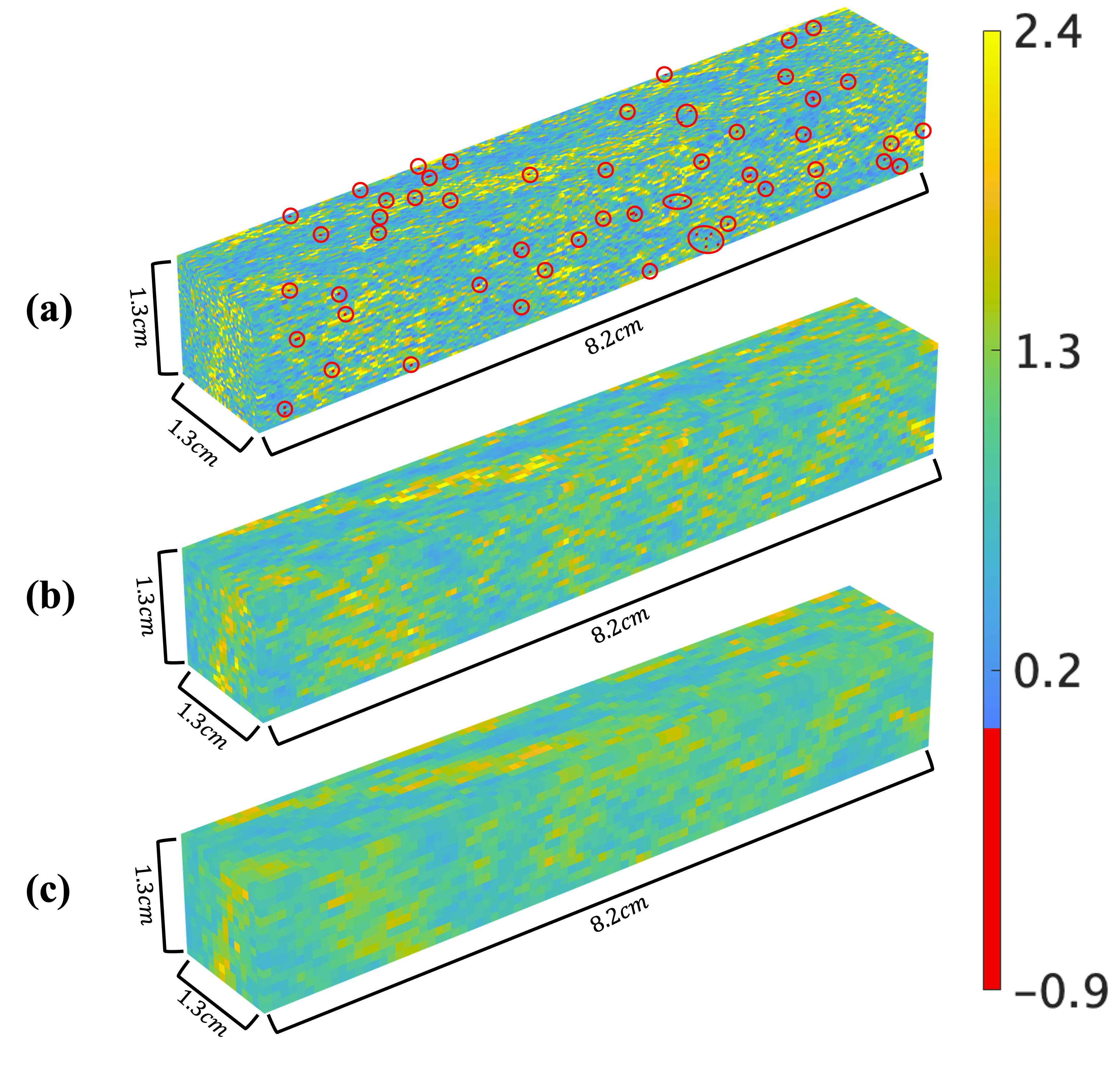}
\caption{Dimensionless velocity distributions normalized by the mean flow velocity of the whole core obtained from the Darcy flow solver: (a)Case I, (b) Case II, (c) Case III. } 
\label{fig:FluxMap}
\end{figure}

Figure \ref{fig:FluxMap} displays the dimensionless velocity (normalized by the mean flow velocity) distributions within the core, for three cases with varying subvolume sizes. In Case I, significant localized velocity variations were evident, with regions deviating both positively and negatively (negative area highlighted by the red circle in \ref{fig:FluxMap}(a)), highlighting the fine-scale heterogeneities captured at this resolution. Negative velocities are observed exclusively in Case I, as many subvolumes are predominantly composed of either flow paths or solid. 
This composition results in notable local permeability variations, leading to occurrences of backflow (Figure \ref{fig:P_K_P_xplot}(a)). Case II, showed a moderately smoothed velocity distribution, where localized variations are still apparent but less pronounced due to intermediate averaging effects.  Case III presented the smoothest velocity field, with evenly distributed variations and minimal representation of fine-scale heterogeneities due to averaging effects.

\begin{figure}[t]
\includegraphics[width=\columnwidth]{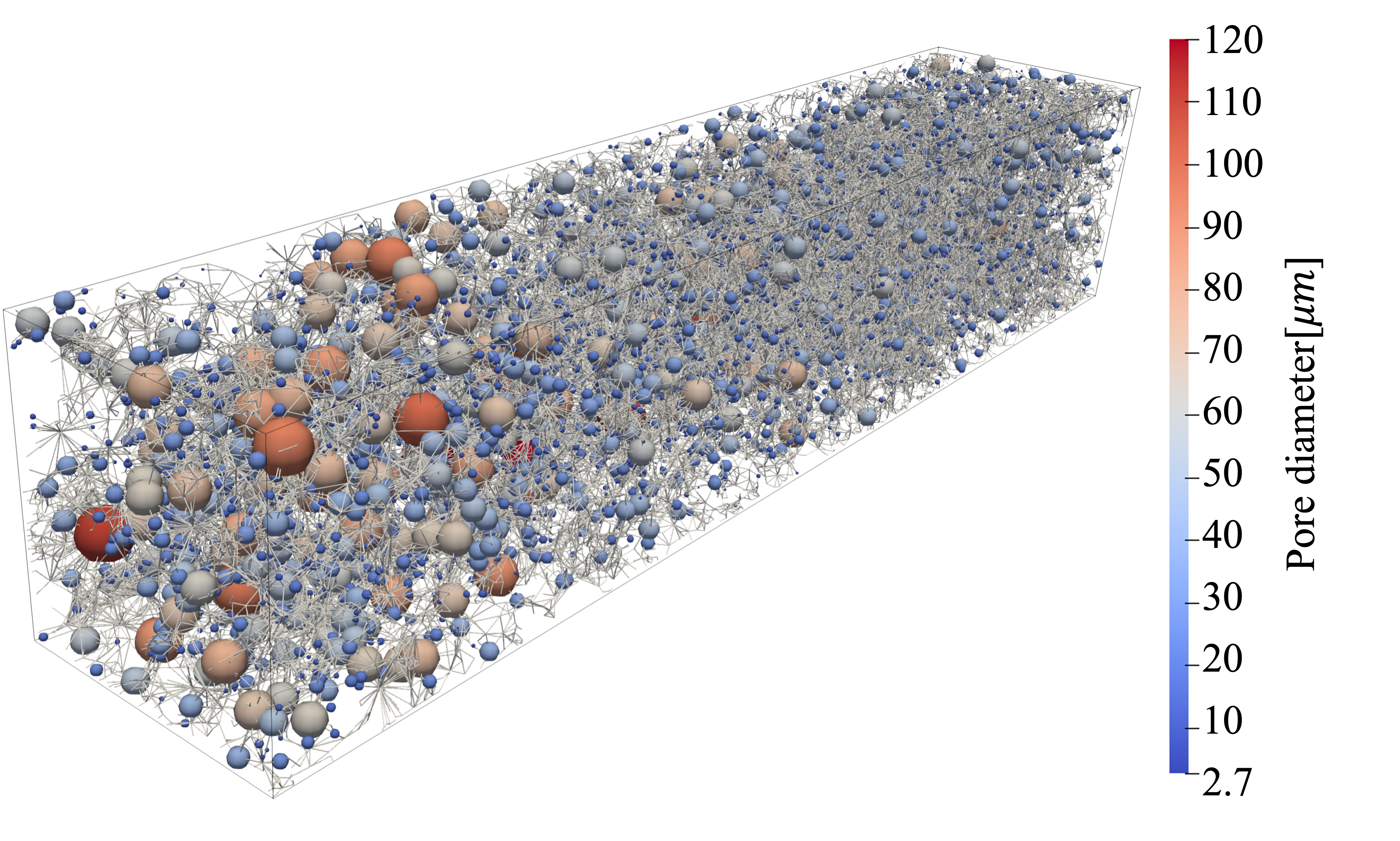}
\caption{Three-dimensional visualization of the pore network model (PNM) reconstructed from the high-resolution CT image. Spheres represent pore bodies, colored according to their diameter. Throats are shown as connecting lines between pores.}
\label{fig:pored}
\end{figure}

The upscaled permeability is determined by analyzing the flow flux distribution within the computational domain using the Darcy flow solver. For comparison, the arithmetic average permeability was also directly calculated from the subvolumes. Additionally, pore network modeling (PNM) was employed to estimate the permeability of the central core areas. PNM simplifies the complex microstructure of porous media by representing the pore space as a network of interconnected nodes (pores) and edges (throats). By solving transport equations based on idealized pore geometries (Figure \ref{fig:pored}), PNM provides an efficient numerical estimation of flow characteristics \citep{blunt2001flow}. The network construction relies on extracting pore geometries from binarized rock CT images with high resolution. In this study, the pore network model was reconstructed using the SNOW algorithm \citep{gostick2019porespy}, and fluid simulations were performed with OpenPNM \citep{gostick2016openpnm}. For the PNM calculations, water was chosen as the working fluid, and a pressure gradient boundary condition was applied along the flow direction to drive fluid movement. The large-scale permeability results obtained from CNN, PNM, and experimental methods are summarized in Table \ref{table:2}.

\begin{table}[h!]
\centering
\begin{tabular}{cccc}
\hline Case & I & II & III \\
\hline K of whole core (upscaled) & 10.9 D & 14.2 D & 12.2 D \\
Arithmetic average permeability & 15.3 D & 16.1 D & 12.9 D \\
K of center core (PNM) & & 19.1 D & \\
Coreflooding experiment data& & 8.9D & \\
\hline
\end{tabular}
\caption{Large-scale permeabilities \(K\) obtained by different methods. (1D $\approx 9.869 \times 10^{-12}$ m$^2$).}
\label{table:2}
\end{table}

The permeability of the entire core, measured through core flooding experiments, was determined to be 8.9 Darcy (D). 
In contrast, the permeability estimated using PNM was significantly higher 
because PNM models the pore space using simplified geometries, which fail to adequately represent the intricate pore structures of real rock samples. 
Improper definition of boundaries between pores and throats can introduce substantial errors in the prediction of transport properties \citep{liu2022pore}. This limitation is particularly pronounced in rocks with small pore sizes, such as Boise sandstone studied here, where these oversimplifications can misrepresent pore connectivity and flow behavior, leading to an underestimation of fluid resistance and an overestimation of permeability. In contrast, our method integrates high-resolution flow data with the raw low-resolution CT scans, preserving the gray-scale information that reflects pore structure, resulting in significantly improved accuracy in permeability predictions (Table \ref{table:2}).
In terms of computation efficiency, the PNM still requires several hours of computation. In contrast,  our deep-learning-based method needs only about 40 mininutes to obtain the upscaled permeability data for Case III (38 minutes for generating the permeability map and 2 minutes for performing the Darcy flow simulation). Thus, our method offers clear advantages in terms of computational time, spatial coverage, and prediction reliability.

A comparison between the upscaled permeability calculated using the Darcy flow solver and the arithmetic average permeability (Table \ref{table:2}) reveals that the discrepancy diminishes as the subvolume size increases. As the subvolume size increases, the local heterogeneity is progressively averaged out. The decreased variability in permeability observed in the large subvolume case results in a more uniformly distributed flux during Darcy flow simulations (Figure \ref{fig:FluxMap}(c)). This uniformity reduces the disparity between the upscaled permeability and the arithmetic average permeability.

The upscaled permeability of 10.9D in Case I initially appears to align most closely with the values obtained from the flooding experiments. However, as illustrated in Figure \ref{fig:Permeability_Porosity_R2}(a), the prediction accuracy of the neural network in Case I is the lowest, with significant deviations observed between the predicted and actual permeability values. Furthermore, the prediction plot for Case I (test dataset) in Figure \ref{fig:AI-Permeability-plot} reveals that the predicted permeability values predominantly fall below the red line, indicating a systematic underestimation of permeability for unknown subvolumes by the trained network. 
This suggests that the impact of local heterogeneity on Darcy flow simulations is most pronounced in this case. 
Such an overrepresentation of heterogeneity could introduce artifacts analogous to noise in the simulation.
Such limitations likely contributed to the observed underestimation of permeability and the reduced prediction accuracy. Consequently, the upscaled permeability calculated for Case I is considered unreliable.
In contrast, Cases II and III demonstrate higher prediction accuracy of the neural network (Figure \ref{fig:Permeability_Porosity_R2}(a) and Figure \ref{fig:AI-Permeability-plot}),
which can improve the overall reliability of the upscaling process.

The permeability distribution histograms (Figure \ref{fig:histgramC}) reveal that many subvolumes with high permeability are present in Case II, which can reconnect and function as flow paths, resulting in an overestimation of the upscaled permeability. In contrast, Case III yields a permeability value of 12.2D from the Darcy flow simulation, which aligns more closely with the experimental result compared to Case II. In this case, the larger subvolumes effectively average out unnecessary local heterogeneities, thereby minimizing their impact on the Darcy flow simulation. Additionally, the neural network achieves the highest prediction accuracy in Case III, allowing for a more reliable evaluation of the upscaled permeability. 
Therefore, among the tested cases, Case III provided the closest agreement with experimental results and the highest prediction accuracy, suggesting its suitability under the current conditions.

\section{Summary}

This study developed a deep learning-based workflow for predicting the upscaled absolute permeability of large rock cores using low-resolution (20$\mu$m/voxel) CT images. 

A clear trend was observed in which increasing the subvolume size led to improved prediction accuracy, reduced data variability, and better agreement with experimental results.
Within the tested range of subvolume sizes ($100^3$–$300^3$), the $300^3$ size exhibited the best overall performance in terms of prediction accuracy and computational stability, and the following main conclusions are drawn:
\begin{enumerate}
 \item Compared with the traditional Kozeny-Carman (KC) model, the machine learning-based approach effectively captured the permeability-porosity relationship by reproducing not only the overall trend of the distribution but also the spread and shape of the data. This capability highlights the ability of machine learning to account for pore microstructure heterogeneity, offering more accurate and reliable predictions compared to traditional empirical models.

 \item The predicted permeability map of the entire core sample demonstrated strong spatial alignment with the rock structure observed in raw CT images. Permeability variations corresponded closely to the density of CT images, with less dense regions showing higher permeability and denser regions exhibiting lower permeability. This spatial alignment validates the accuracy of the machine learning-based predictions. 

 \item Using the predicted permeability maps, the study successfully calculated upscaled permeability with higher accuracy than traditional PNM because our proposed method integrates high-resolution flow data with low-resolution CT scans, preserving detailed pore structure information and enabling more accurate and reliable predictions of permeability at the core scale. 

 \item Finally, the workflow demonstrated its ability to upscale permeability from the millimeter-scale sub-core to the centimeter-scale core sample, effectively integrating small-scale heterogeneity into larger-scale models. These findings underscore the robustness and scalability of the machine learning-based workflow for simulating fluid flow in porous media across multiple scales, providing a powerful framework to advance research in digital rock physics and geosciences. 
\end{enumerate}

While the results are promising, several limitations must be acknowledged. 
Within the tested range of subvolume sizes ($100^3$–$300^3$), the subvolume size of $300^3$ (containing approximately 1240 pore bodies) showed the best performance for the Boise sandstone sample, and its generalizability to other rock types or flow scenarios remains untested.

Furthermore, for high-density rocks such as granite, basalt, or quartzite, where pore sizes are extremely small and often disconnected, it may be challenging to capture voids accurately enough to calculate physical properties using CT images. This limitation suggests that the current workflow may struggle with rocks exhibiting extremely low porosity, such as those typically found in crystalline or metamorphic formations. 
Future research should focus on extending the workflow to diverse rock types and fluid conditions to evaluate its general applicability. Additionally, the workflow could be expanded to predict other rock properties beyond permeability, such as electrical conductivity, elastic modulus, thermal conductivity, and seismic wave velocities, enabling a more comprehensive characterization of rock samples. Finally, integrating additional data sources along with the CT images, such as mineralogical or chemical properties, could increase the model's versatility and predictive power.





\section*{CRediT authorship contribution statement}
\textbf{Yaotian Guo}: Conceptualization, Methodology, Formal analysis, Investigation, Material preparation, Data curation, Writing – original draft, Writing – review \& editing.  
\textbf{Fei Jiang}: Conceptualization, Methodology, Investigation, Material preparation, Data curation, Writing – review \& editing.  
\textbf{Takeshi Tsuji}: Conceptualization, Methodology, Investigation, Material preparation, Data curation, Funding acquisition, Writing – review \& editing.  
\textbf{Yoshitake Kato}: Investigation, Material preparation, Data curation, Funding acquisition, Writing – review \& editing.  
\textbf{Mai Shimokawara}: Investigation, Material preparation, Data curation, Writing – review \& editing.  
\textbf{Lionel Esteban}: Investigation, Material preparation, Data curation, Writing – review \& editing.  
\textbf{Mojtaba Seyyedi}: Investigation, Material preparation, Data curation, Writing – review \& editing.  
\textbf{Marina Pervukhina}: Investigation, Material preparation, Data curation, Funding acquisition, Writing – review \& editing.  
\textbf{Maxim Lebedev}: Investigation, Material preparation, Data curation, Writing – review \& editing.  
\textbf{Ryuta Kitamura}: Funding acquisition, Writing – review \& editing.  


\section*{Declaration of Competing Interest}
The authors declare that they have no known competing financial
interests or personal relationships that could have appeared to influence
the work reported in this paper.

\section*{Acknowledgements}
We gratefully acknowledge JOGMEC for providing the rock CT images used in this study. This work was also partially supported by JSPS KAKENHI Grant Numbers JP22K03927, JP24H00440 and JP21H05202, and JST SPRING Grant Number JPMJSP2111.



\begin{thebibliography}{00}


\bibitem[Abadi et al.(2015)]{abadi2015tensorflow}
\href{https://doi.org/10.48550/arXiv.1603.04467}{%
Abadi, M., Agarwal, A., Barham, P., Brevdo, E., Chen, Z., Citro, C., Corrado, G.S., Davis, A., Dean, J., Devin, M., et al., 2015.
TensorFlow: Large-scale machine learning on heterogeneous systems.
arXiv preprint arXiv:1603.04467.
https://doi.org/10.48550/arXiv.1603.04467.}

\bibitem[Akhtar et al.(2020)]{akhtar2020characteristics}
\href{https://doi.org/10.5772/intechopen.93800}{%
Akhtar, N., Syakir, M.I., Anees, M.T., Qadir, A., Yusuff, M.S., 2020.
Characteristics and assessment of groundwater.
Groundwater Management and Resources. IntechOpen, London.
https://doi.org/10.5772/intechopen.93800.}

\bibitem[Bachu(2008)]{bachu2008ccs}
\href{https://doi.org/10.1016/j.pecs.2007.10.001}{%
Bachu, S., 2008.
CO$_2$ storage in geological media: Role, means, status and barriers to deployment.
Prog. Energy Combust. Sci. 34(2), 254--273.
https://doi.org/10.1016/j.pecs.2007.10.001.}

\bibitem[Bade et al.(2024)]{bade2024hydrogen}
\href{https://doi.org/10.1016/j.ijhydene.2024.07.187}{%
Bade, S.O., Taiwo, K., Ndulue, U.F., Tomomewo, O.S., Oni, B.A., 2024.
A review of underground hydrogen storage systems: Current status, modeling approaches, challenges, and future prospective.
Int. J. Hydrogen Energy 80, 449--474.
https://doi.org/10.1016/j.ijhydene.2024.07.187.}

\bibitem[Bear(2013)]{bear2013dynamics}
Bear, J., 2013.
Dynamics of fluids in porous media.
Courier Corporation, Mineola, NY.

\bibitem[Blunt(2001)]{blunt2001flow}
\href{https://doi.org/10.1016/S1359-0294(01)00084-X}{%
Blunt, M.J., 2001.
Flow in porous media—pore-network models and multiphase flow.
Curr. Opin. Colloid Interface Sci. 6(3), 197--207.
https://doi.org/10.1016/S1359-0294(01)00084-X.}

\bibitem[Carman(1937)]{carman1937fluid}
\href{https://doi.org/10.1016/S0263-8762(97)80003-2}{%
Carman, P.C., 1937.
Fluid flow through granular beds.
Trans. Inst. Chem. Eng. 15, 150--166.
https://doi.org/10.1016/S0263-8762(97)80003-2.}

\bibitem[Chen et al.(2023)]{chen2023predicting}
\href{https://doi.org/10.1029/2023WR035521}{%
Chen, X., Yang, J., Ma, L., Rabbani, A., Babaei, M., 2023.
Predicting 3D physical properties from a single 2D slice based on convolutional neural networks.
Water Resour. Res. 59(9), e2023WR035521.
https://doi.org/10.1029/2023WR035521.}

\bibitem[Cheung et al.(2012)]{cheung2012effect}
\href{https://doi.org/10.1029/2012GL053739}{%
Cheung, C.S.N., Baud, P., Wong, T.-f., 2012.
Effect of grain size distribution on the development of compaction localization in porous sandstone.
Geophys. Res. Lett. 39(21).
https://doi.org/10.1029/2012GL053739.}

\bibitem[Clauser(1992)]{clauser1992permeability}
\href{https://doi.org/10.1029/91EO00190}{%
Clauser, C., 1992.
Permeability of crystalline rocks.
Eos Trans. AGU 73(21), 233--238.
https://doi.org/10.1029/91EO00190.}

\bibitem[Cnudde and Boone(2013)]{cnudde2013high}
\href{https://doi.org/10.1016/j.earscirev.2013.04.003}{%
Cnudde, V., Boone, M.N., 2013.
High-resolution X-ray computed tomography in geosciences.
Earth-Sci. Rev. 123, 1--17.
https://doi.org/10.1016/j.earscirev.2013.04.003.}

\bibitem[Costa(2006)]{costa2006permeability}
\href{https://doi.org/10.1029/2005GL025134}{%
Costa, A., 2006.
Permeability-porosity relationship: A reexamination of the Kozeny-Carman equation.
Geophys. Res. Lett. 33(2).
https://doi.org/10.1029/2005GL025134.}

\bibitem[Draper and Smith(1998)]{draper1998applied}
\href{https://doi.org/10.1002/9781118625590}{%
Draper, N.R., Smith, H., 1998.
Applied regression analysis.
Wiley, New York.
https://doi.org/10.1002/9781118625590.}

\bibitem[d'Humieres(2002)]{dhumieres2002multiple}
\href{https://doi.org/10.1098/rsta.2001.0955}{%
d'Humi\`eres, D., 2002.
Multiple--relaxation--time lattice Boltzmann models in three dimensions.
Philos. Trans. R. Soc. Lond. A 360(1792), 437--451.
https://doi.org/10.1098/rsta.2001.0955.}

\bibitem[Durlofsky(2005)]{durlofsky2005upscaling}
Durlofsky, L.J., 2005.
Upscaling and gridding of fine scale geological models for flow simulation.
Comput. Geosci. 29, 399--419.

\bibitem[Elmorsy et al.(2023)]{elmorsy2023rapid}
\href{https://doi.org/10.1029/2023WR035064}{%
Elmorsy, M., El-Dakhakhni, W., Zhao, B., 2023.
Rapid permeability upscaling of digital porous media via physics-informed neural networks.
Water Resour. Res. 59(12), e2023WR035064.
https://doi.org/10.1029/2023WR035064.}

\bibitem[Erfani et al.(2024)]{erfani2024spatial}
\href{https://doi.org/10.1007/s11242-023-02044-x}{%
Erfani, H., Haghani, R., McClure, J., Boek, E., Berg, C.F., 2024.
Spatial characterization of wetting in porous media using local lattice-Boltzmann simulations.
Transp. Porous Media 151(3), 429--448.
https://doi.org/10.1007/s11242-023-02044-x.}

\bibitem[Esmaeilpour et al.(2021)]{esmaeilpour2021scale}
\href{https://doi.org/10.1016/j.fuel.2021.121090}{%
Esmaeilpour, M., Ghanbarian, B., Liang, F., Liu, H.-H., 2021.
Scale-dependent permeability and formation factor in porous media: Applications of percolation theory.
Fuel 301, 121090.
https://doi.org/10.1016/j.fuel.2021.121090.}

\bibitem[Esmaeilpour et al.(2023)]{esmaeilpour2023estimating}
\href{https://doi.org/10.1029/2022WR033462}{%
Esmaeilpour, M., Ghanbarian, B., Sousa, R., King, P.R., 2023.
Estimating permeability and its scale dependence using renormalization group theory.
Water Resour. Res. 59(5), e2022WR033462.
https://doi.org/10.1029/2022WR033462.}

\bibitem[Gostick et al.(2016)]{gostick2016openpnm}
\href{https://doi.org/10.1109/MCSE.2016.49}{%
Gostick, J., Aghighi, M., Hinebaugh, J., Tranter, T., Hoeh, M.A., Day, H., Spellacy, B., Sharqawy, M.H., Bazylak, A., Burns, A., et al., 2016.
OpenPNM: A pore network modeling package.
Comput. Sci. Eng. 18(4), 60--74.
https://doi.org/10.1109/MCSE.2016.49.}

\bibitem[Gostick et al.(2019)]{gostick2019porespy}
\href{https://doi.org/10.21105/joss.01296}{%
Gostick, J.T., Khan, Z.A., Tranter, T.G., Kok, M.D.R., Agnaou, M., Sadeghi, M., Jervis, R., 2019.
PoreSpy: A python toolkit for quantitative analysis of porous media images.
J. Open Source Softw. 4(37), 1296.
https://doi.org/10.21105/joss.01296.}

\bibitem[Hayatdavoudi et al.(2025)]{hayatdavoudi2025comparative}
\href{https://doi.org/10.1038/s41598-025-07211-2}{%
Hayatdavoudi, M., Niri, M.E., Kalhor, A., 2025.
Comparative analysis of sandstone microtomographic image segmentation using advanced CNNs.
Sci. Rep. 15(1), 22164.
https://doi.org/10.1038/s41598-025-07211-2.}

\bibitem[He and Luo(1997)]{he1997theory}
\href{https://doi.org/10.1103/PhysRevE.56.6811}{%
He, X., Luo, L.-S., 1997.
Theory of the lattice Boltzmann method: From the Boltzmann equation to the lattice Boltzmann equation.
Phys. Rev. E 56(6), 6811.
https://doi.org/10.1103/PhysRevE.56.6811.}

\bibitem[Huang et al.(2022)]{huang2022deep}
\href{https://doi.org/10.1103/PhysRevE.105.015308}{%
Huang, Y., Xiang, Z., Qian, M., 2022.
Deep-learning-based porous media microstructure quantitative characterization and reconstruction method.
Phys. Rev. E 105(1), 015308.
https://doi.org/10.1103/PhysRevE.105.015308.}

\bibitem[Jiang and Tsuji(2014)]{jiang2014changes}
\href{https://doi.org/10.1103/PhysRevE.90.053306}{%
Jiang, F., Tsuji, T., 2014.
Changes in pore geometry and relative permeability caused by carbonate precipitation in porous media.
Phys. Rev. E 90(5), 053306.
https://doi.org/10.1103/PhysRevE.90.053306.}

\bibitem[Jiang and Tsuji(2015)]{jiang2015impact}
\href{https://doi.org/10.1002/2014WR016070}{%
Jiang, F., Tsuji, T., 2015.
Impact of interfacial tension on residual CO$_2$ clusters in porous sandstone.
Water Resour. Res. 51(3), 1710--1722.
https://doi.org/10.1002/2014WR016070.}

\bibitem[Jiang and Tsuji(2017)]{jiang2017estimation}
\href{https://doi.org/10.1002/2016WR019098}{%
Jiang, F., Tsuji, T., 2017.
Estimation of three-phase relative permeability by simulating fluid dynamics directly on rock-microstructure images.
Water Resour. Res. 53(1), 11--32.
https://doi.org/10.1002/2016WR019098.}

\bibitem[Jiang et al.(2023)]{jiang2023upscaling}
\href{https://doi.org/10.1029/2022WR033267}{%
Jiang, F., Guo, Y., Tsuji, T., Kato, Y., Shimokawara, M., Esteban, L., Seyyedi, M., Pervukhina, M., Lebedev, M., Kitamura, R., 2023.
Upscaling permeability using multiscale X-ray-CT images with digital rock modeling and deep learning techniques.
Water Resour. Res. 59(3), e2022WR033267.
https://doi.org/10.1029/2022WR033267.}

\bibitem[Jiang et al.(2024)]{jiang2024svm}
\href{https://doi.org/10.1016/j.advwatres.2023.104605}{%
Jiang, H., Arns, C., Yuan, Y., Qin, C.-Z., 2024.
SVM-based fast 3D pore-scale rock-typing and permeability upscaling for complex rocks using Minkowski functionals.
Adv. Water Resour. 183, 104605.
https://doi.org/10.1016/j.advwatres.2023.104605.}

\bibitem[Kalule et al.(2024)]{kalule4781996relative}
\href{https://doi.org/10.2139/ssrn.4781996}{%
Kalule, R., Abderrahmane, H., Ahmed, S., Alameri, W., Sassi, M., 2024.
Relative permeability and capillary pressure estimation via physics-informed machine learning and reinforcement learning.
SSRN 4781996.
https://doi.org/10.2139/ssrn.4781996.}

\bibitem[Kang et al.(2024)]{kang2024hybrid}
\href{https://doi.org/10.1016/j.compgeo.2024.106163}{%
Kang, Q., Li, K.-Q., Fu, J.-L., Liu, Y., 2024.
Hybrid LBM and machine learning algorithms for permeability prediction of porous media: A comparative study.
Comput. Geotech. 168, 106163.
https://doi.org/10.1016/j.compgeo.2024.106163.}

\bibitem[Khan and Khanal(2024)]{khan2024machine}
\href{https://doi.org/10.1021/acsomega.3c10131}{%
Khan, M.I., Khanal, A., 2024.
Machine learning assisted prediction of porosity and related properties using digital rock images.
ACS Omega 9(28), 30205--30223.
https://doi.org/10.1021/acsomega.3c10131.}

\bibitem[Ko et al.(2023)]{ko2023prediction}
\href{https://doi.org/10.1007/s11242-023-01961-1}{%
Ko, D.D., Ji, H., Ju, Y.S., 2023.
Prediction of 3D velocity field of reticulated foams using deep learning for transport analysis.
Transp. Porous Media 148(3), 577--604.
https://doi.org/10.1007/s11242-023-01961-1.}

\bibitem[Koroteev and Tekic(2021)]{koroteev2021artificial}
\href{https://doi.org/10.1016/j.egyai.2020.100041}{%
Koroteev, D., Tekic, Z., 2021.
Artificial intelligence in oil and gas upstream: Trends, challenges, and scenarios for the future.
Energy AI 3, 100041.
https://doi.org/10.1016/j.egyai.2020.100041.}

\bibitem[Krizhevsky et al.(2012)]{krizhevsky2012imagenet}
\href{https://doi.org/10.1145/3065386}{%
Krizhevsky, A., Sutskever, I., Hinton, G.E., 2012.
Imagenet classification with deep convolutional neural networks.
Adv. Neural Inf. Process. Syst. 25.
https://doi.org/10.1145/3065386.}

\bibitem[LeCun et al.(1998)]{lecun1998gradient}
\href{https://doi.org/10.1109/5.726791}{%
LeCun, Y., Bottou, L., Bengio, Y., Haffner, P., 1998.
Gradient-based learning applied to document recognition.
Proc. IEEE 86(11), 2278--2324.
https://doi.org/10.1109/5.726791.}

\bibitem[LeCun et al.(2015)]{lecun2015deep}
\href{https://doi.org/10.1038/nature14539}{%
LeCun, Y., Bengio, Y., Hinton, G., 2015.
Deep learning.
Nature 521(7553), 436--444.
https://doi.org/10.1038/nature14539.}

\bibitem[Li and Xie(2021)]{li2021calculation}
\href{https://doi.org/10.1093/jge/gxab065}{%
Li, G., Xie, C., 2021.
Calculation of matrix permeability from velocity and attenuation of ultrasonic S-wave.
J. Geophys. Eng. 18(6), 984--994.
https://doi.org/10.1093/jge/gxab065.}

\bibitem[Li et al.(2016)]{li2016lattice}
\href{https://doi.org/10.48550/arXiv.1609.09685}{%
Li, J., Ho, M.T., Wu, L., Zhang, Y., 2016.
Lattice Boltzmann modelling of intrinsic permeability.
arXiv preprint arXiv:1609.09685.
https://doi.org/10.48550/arXiv.1609.09685.}

\bibitem[Lie(2019)]{lie2019introduction}
\href{https://doi.org/10.1017/9781108591416}{%
Lie, K.-A., 2019.
An introduction to reservoir simulation using MATLAB/GNU Octave: User guide for the MATLAB reservoir simulation toolbox (MRST).
Cambridge University Press, Cambridge.
https://doi.org/10.1017/9781108591416.}

\bibitem[Liu et al.(2022)]{liu2022pore}
\href{https://doi.org/10.1029/2022WR033142}{%
Liu, Y., Gong, W., Zhao, Y., Jin, X., Wang, M., 2022.
A pore-throat segmentation method based on local hydraulic resistance equivalence for pore-network modeling.
Water Resour. Res. 58(12), e2022WR033142.
https://doi.org/10.1029/2022WR033142.}

\bibitem[Louis et al.(2024)]{louis2024role}
\href{https://doi.org/10.1007/s00603-024-04152-6}{%
Louis, L., Boyd, P., Hofmann, R., Saxena, N., 2024.
On the role and evolution of local grain size heterogeneity during confined compression of Boise sandstone as seen by X-ray micro-CT imaging.
Rock Mech. Rock Eng. 1--22.
https://doi.org/10.1007/s00603-024-04152-6.}

\bibitem[Massarweh and Abushaikha(2022)]{massarweh2022review}
\href{https://doi.org/10.1016/j.petlm.2021.05.002}{%
Massarweh, O., Abushaikha, A.S., 2022.
A review of recent developments in CO$_2$ mobility control in enhanced oil recovery.
Petroleum 8(3), 291--317.
https://doi.org/10.1016/j.petlm.2021.05.002.}

\bibitem[Mishra et al.(2024)]{mishra2024pore}
\href{https://doi.org/10.1016/j.acags.2024.100179}{%
Mishra, A., Ma, L., Reddy, S.C., Attanayake, J., Haese, R.R., 2024.
Pore-to-Darcy scale permeability upscaling for media with dynamic pore structure using graph theory.
Appl. Comput. Geosci. 100179.
https://doi.org/10.1016/j.acags.2024.100179.}

\bibitem[Onimisi et al.(2023)]{onimisi2023constrained}
\href{https://doi.org/10.1002/fld.5171}{%
Onimisi, T.A., Lashore, B.O., Akanji, L.T., Gomes, J.L.M.A., 2023.
A constrained proper orthogonal decomposition model for upscaling permeability.
Int. J. Numer. Methods Fluids 95(6), 899--916.
https://doi.org/10.1002/fld.5171.}

\bibitem[Qian et al.(1992)]{qian1992lattice}
\href{https://doi.org/10.1209/0295-5075/17/6/001}{%
Qian, Y.-H., d'Humi\`eres, D., Lallemand, P., 1992.
Lattice BGK models for Navier-Stokes equation.
Europhys. Lett. 17(6), 479.
https://doi.org/10.1209/0295-5075/17/6/001.}

\bibitem[Raza et al.(2019)]{raza2019significant}
\href{https://doi.org/10.1016/j.petlm.2018.12.007}{%
Raza, A., Gholami, R., Rezaee, R., Rasouli, V., Rabiei, M., 2019.
Significant aspects of carbon capture and storage--A review.
Petroleum 5(4), 335--340.
https://doi.org/10.1016/j.petlm.2018.12.007.}

\bibitem[Ren et al.(2021)]{ren2021microplastics}
\href{https://doi.org/10.1016/j.jhazmat.2021.126455}{%
Ren, Z., Gui, X., Xu, X., Zhao, L., Qiu, H., Cao, X., 2021.
Microplastics in the soil-groundwater environment: Aging, migration, and co-transport of contaminants--a critical review.
J. Hazard. Mater. 419, 126455.
https://doi.org/10.1016/j.jhazmat.2021.126455.}

\bibitem[Reboulet and Barrash(2003)]{reboulet2003core}
Reboulet, E.C., Barrash, W., 2003.
Core, grain-size, and porosity data from the Boise Hydrogeophysical Research Site, Boise, Idaho.
Center for Geophysical Investigation of the Shallow Subsurface, Technical Report 03--02.

\bibitem[Roslin et al.(2022)]{roslin2022processing}
\href{https://doi.org/10.1016/j.mineng.2022.107748}{%
Roslin, A., Marsh, M., Piché, N., Provencher, B., Mitchell, T.R., Onederra, I.A., Leonardi, C.R., 2022.
Processing of micro-CT images of granodiorite rock samples using CNN, Part I: Super-resolution enhancement using a 3D CNN.
Miner. Eng. 188, 107748.
https://doi.org/10.1016/j.mineng.2022.107748.}

\bibitem[Rybach(2010)]{rybach2010future}
Rybach, L., 2010.
The future of geothermal energy and its challenges.
Proc. World Geothermal Congr., 29.

\bibitem[Simonyan and Zisserman(2014)]{simonyan2014very}
\href{https://doi.org/10.48550/arXiv.1409.1556}{%
Simonyan, K., Zisserman, A., 2014.
Very deep convolutional networks for large-scale image recognition.
arXiv preprint arXiv:1409.1556.
https://doi.org/10.48550/arXiv.1409.1556.}

\bibitem[Suss et al.(2023)]{suss2023hybrid}
\href{https://doi.org/10.2514/6.2023-3433}{%
Suss, A., Mary, I., Le Garrec, T., Mari\'e, S., 2023.
A hybrid lattice Boltzmann-Navier-Stokes method on overset grids.
AIAA Aviation 2023 Forum, 3433.
https://doi.org/10.2514/6.2023-3433.}

\bibitem[Szegedy et al.(2015)]{szegedy2015going}
\href{https://doi.org/10.1109/CVPR.2015.7298594}{%
Szegedy, C., Liu, W., Jia, Y., Sermanet, P., Reed, S., Anguelov, D., Erhan, D., Vanhoucke, V., Rabinovich, A., 2015.
Going deeper with convolutions.
Proc. IEEE Conf. Comput. Vis. Pattern Recognit., 1--9.
https://doi.org/10.1109/CVPR.2015.7298594.}

\bibitem[Taiwo et al.(2024)]{taiwo2024storage}
\href{https://doi.org/10.1016/j.est.2024.111844}{%
Taiwo, G.O., Tomomewo, O.S., Oni, B.A., 2024.
A comprehensive review of underground hydrogen storage: Insight into geological sites, economics, barriers, and outlook.
J. Energy Storage 90, 111844.
https://doi.org/10.1016/j.est.2024.111844.}

\bibitem[Telvari et al.(2023)]{telvari2023prediction}
\href{https://doi.org/10.1016/j.advwatres.2023.104442}{%
Telvari, S., Sayyafzadeh, M., Siavashi, J., Sharifi, M., 2023.
Prediction of two-phase flow properties for digital sandstones using 3D convolutional neural networks.
Adv. Water Resour. 176, 104442.
https://doi.org/10.1016/j.advwatres.2023.104442.}

\bibitem[Tripathi et al.(2018)]{tripathi2018exporting}
Tripathi, D.N., Hathon, L.A., Myers, M.T., 2018.
Exporting petrophysical properties of sandstones from thin section image analysis.
SPWLA Annual Logging Symp., D043S006R001.

\bibitem[Tsuji et al.(2016)]{tsuji2016characterization}
\href{https://doi.org/10.1016/j.advwatres.2016.03.005}{%
Tsuji, T., Jiang, F., Christensen, K.T., 2016.
Characterization of immiscible fluid displacement processes with various capillary numbers and viscosity ratios in 3D natural sandstone.
Adv. Water Resour. 95, 3--15.
https://doi.org/10.1016/j.advwatres.2016.03.005.}

\bibitem[Wang and Zai(2023)]{wang2023image}
\href{https://doi.org/10.1016/j.advwatres.2023.104384}{%
Wang, F., Zai, Y., 2023.
Image segmentation and flow prediction of digital rock with U-net network.
Adv. Water Resour. 172, 104384.
https://doi.org/10.1016/j.advwatres.2023.104384.}

\bibitem[Wang et al.(2021)]{wang2021deep}
\href{https://doi.org/10.1016/j.earscirev.2021.103555}{%
Wang, Y., Blunt, M.J., Armstrong, R.T., Mostaghimi, P., 2021.
Deep learning in pore scale imaging and modeling.
Earth-Sci. Rev. 215, 103555.
https://doi.org/10.1016/j.earscirev.2021.103555.}

\bibitem[Wang et al.(2023a)]{wang2023pore}
\href{https://doi.org/10.1016/j.earscirev.2023.104602}{%
Wang, W., Xie, Q., An, S., Bakhshian, S., Kang, Q., Wang, H., Xu, X., Su, Y., Cai, J., Yuan, B., 2023.
Pore-scale simulation of multiphase flow and reactive transport processes involved in geologic carbon sequestration.
Earth-Sci. Rev. 104602.
https://doi.org/10.1016/j.earscirev.2023.104602.}

\bibitem[Wang et al.(2023b)]{wang2023recent}
\href{https://doi.org/10.1177/09544062221077583}{%
Wang, L., Liu, Z., Rajamuni, M., 2023.
Recent progress of lattice Boltzmann method and its applications in fluid-structure interaction.
Proc. IMechE C J. Mech. Eng. Sci. 237(11), 2461--2484.
https://doi.org/10.1177/09544062221077583.}

\bibitem[Withers et al.(2021)]{withers2021ct}
\href{https://doi.org/10.1038/s43586-021-00015-4}{%
Withers, P.J., Bouman, C., Carmignato, S., Cnudde, V., Grimaldi, D., Helfen, L., Stock, S.R., 2021.
X-ray computed tomography.
Nat. Rev. Methods Primers 1(1), 1--21.
https://doi.org/10.1038/s43586-021-00015-4.}

\bibitem[Wu et al.(2024)]{wu2024geothermal}
\href{https://doi.org/10.1007/s11053-024-10361-1}{%
Wu, J., Xu, H., Xiong, B., Fang, C., Wang, S., Zong, P., Liu, D., 2024.
Effects of mineral displacement on geothermal reservoir properties at high temperatures identified using micro-CT and digital volume correlation.
Nat. Resour. Res. 33, 1613--1623.
https://doi.org/10.1007/s11053-024-10361-1.}

\bibitem[Xie et al.(2023)]{xie2023direct}
\href{https://doi.org/10.1016/j.ijrmms.2023.105544}{%
Xie, C., Zhu, J., Wang, J., Yang, J., Song, H., 2023.
Direct prediction of relative permeability curve from 3D digital rock images based on deep learning approaches.
Int. J. Rock Mech. Min. Sci. 170, 105544.
https://doi.org/10.1016/j.ijrmms.2023.105544.}

\bibitem[Yang et al.(2018)]{yang2018flow}
\href{https://doi.org/10.3390/en11082145}{%
Yang, Y., Liu, Z., Yao, J., Zhang, L., Ma, J., Hejazi, S.H., Luquot, L., Ngarta, T.D., 2018.
Flow simulation of artificially induced microfractures using digital rock and lattice Boltzmann methods.
Energies 11(8), 2145.
https://doi.org/10.3390/en11082145.}

\bibitem[Yousefzadeh and Ahmadi(2023)]{yousefzadeh2023fast}
\href{https://doi.org/10.1016/j.geoen.2023.212211}{%
Yousefzadeh, R., Ahmadi, M., 2023.
Fast marching method assisted permeability upscaling using a hybrid deep learning method coupled with particle swarm optimization.
Geoenergy Sci. Eng. 230, 212211.
https://doi.org/10.1016/j.geoen.2023.212211.}

\bibitem[Youssef et al.(2007)]{youssef2007high}
\href{https://doi.org/10.2118/111427-MS}{%
Youssef, S., Rosenberg, E., Gland, N.F., Kenter, J.A., Skalinski, M., Vizika, O., 2007.
High resolution CT and pore-network models to assess petrophysical properties of homogeneous and heterogeneous carbonates.
SPE/EAGE Reservoir Characterization and Simulation Conf.
https://doi.org/10.2118/111427-MS.}

\bibitem[Zhai et al.(2020)]{zhai2020migration}
\href{https://doi.org/10.1016/j.jngse.2020.103233}{%
Zhai, H., Xue, Z., Park, H., Aizawa, Y., Baba, Y., Zhang, Y., 2020.
Migration characteristics of supercritical CO$_2$ microbubble flow in the Berea sandstone revealed by voxel-based X-ray computed tomography imaging analysis.
J. Nat. Gas Sci. Eng. 77, 103233.
https://doi.org/10.1016/j.jngse.2020.103233.}
\end{thebibliography}



\end{document}